\documentclass[aps,prb,twocolumn,color,pdflatex,citeautoscript,superscriptaddress]{revtex4-1}
\usepackage[T1]{fontenc}
\usepackage{graphicx}% Include figure files
\usepackage{dcolumn}% Align table columns on decimal point
\usepackage{amsmath,amssymb,bbold,bm}
\usepackage{hyperref}
\hypersetup{colorlinks=true,citecolor=blue,linkcolor=blue,urlcolor=blue,pdfstartview=FitH,pdfpagemode=PageWidth}
\usepackage{amsfonts}
\usepackage{listings}
\usepackage{braket}
\usepackage{float,latexsym}
\usepackage{multirow}

\usepackage{color}
\definecolor{dkgreen}{rgb}{0,0.5,0}

\newcommand{\comment}[1]{}{}
\begin{document}

\title{Interaction effects in superconductor/quantum spin Hall devices: universal transport signatures and fractional Coulomb blockade}
\author{David Aasen}
\affiliation{Department of Physics and Institute for Quantum Information and Matter, California Institute of Technology, Pasadena, CA 91125, USA}
\author{Shu-Ping Lee}
\affiliation{Department of Physics and Institute for Quantum Information and Matter, California Institute of Technology, Pasadena, CA 91125, USA}
\affiliation{Department of Physics, University of Alberta, Edmonton, Alberta T6G 2E1, Canada}
\author{Torsten Karzig}
\affiliation{Department of Physics and Institute for Quantum Information and Matter, California Institute of Technology, Pasadena, CA 91125, USA}
\affiliation{Station Q, Microsoft Research, Santa Barbara, California 93106-6105, USA}
\author{Jason Alicea}
\affiliation{Department of Physics and Institute for Quantum Information and Matter, California Institute of Technology, Pasadena, CA 91125, USA}
\affiliation{Walter Burke Institute for Theoretical Physics, California Institute of Technology, Pasadena, CA 91125, USA}

\date{\today}

\begin{abstract}
Interfacing $s$-wave superconductors and quantum spin Hall edges produces time-reversal-invariant topological superconductivity of a type that can not arise in strictly 1D systems.  With the aim of establishing sharp fingerprints of this novel phase, we use renormalization group methods to extract universal transport characteristics of superconductor/quantum spin Hall heterostructures where the native edge states serve as leads.  We determine scaling forms for the conductance through a grounded superconductor and show that the results depend sensitively on the interaction strength in the leads, the size of the superconducting region, and the presence or absence of time-reversal-breaking perturbations.  We also study transport across a floating superconducting island isolated by magnetic barriers.  Here we predict $e$-periodic Coulomb-blockade peaks, as recently observed in nanowire devices\cite{Exponential}, with the added feature that the island can support \emph{fractional} charge tunable via the relative orientation of the barrier magnetizations.  As an interesting corollary, when the magnetic barriers arise from strong interactions at the edge that spontaneously break time-reversal symmetry, the Coulomb-blockade periodicity changes from $e$ to $e/2$.  These findings suggest several future experiments that probe unique characteristics of topological superconductivity at the quantum spin Hall edge.  
\end{abstract}

\maketitle

%%%%%%%%%%%%%%%%%%%%%%%%%%%%%%%%%%%%%%%%%%%%%%%%

\section{Introduction}

One-dimensional (1D) topological superconductors\cite{Kitaev-Unpaired-Majorana-1D-wire,BeenakkerReview,aliceaNewDirectionofMajorana,FlensbergReview,FranzReview,NayakReview} present great opportunities both for new physics and longer-term fault-tolerant quantum computing applications\cite{KitaevTQC,NayakQuantumComputation}.  Quantum-spin-Hall (QSH) systems offer a rather unique platform:\cite{Fu-Kane-QSH2}  Coupling the helical edge states to an $s$-wave superconductor naturally generates \emph{time-reversal-invariant} topological superconductivity\cite{Fu-Kane-QSH2} provided the proximitized edge is sufficiently long 
and the adjacent bulk is depleted of carriers.  On general grounds, the topological phase created under these conditions cannot exist in strictly 1D time-reversal-symmetric systems (as created, e.g., in blueprints from Refs.~\onlinecite{JaySau1DSemiconductor,Gil1DSemiconductor,Choy,Yazdani}).\footnote{A similar topological phase in strict 1D would carry an unpaired Majorana zero mode at each boundary, necessitating broken time-reversal symmetry.  QSH systems avoid this obstruction since the edge states always form a closed loop in space.  Qualitatively different time-reversal topological superconductors can, however, appear in strictly 1D systems.  One example is the Kitaev chain with a time-reversal symmetry $\mathcal{T}$ that squares to $+1$; alternatively, strict 1D topological superconductors can enjoy a $\mathcal{T}^2 = -1$ symmetry and support Kramers pairs of Majorana zero modes at each end (see, e.g., Ref.~\onlinecite{RyuClassification}).}  Time-reversal symmetry endows topological superconductivity at the QSH edge with novel and practically useful characteristics, notably resilience\cite{PotterLee} against non-magnetic disorder\footnote{Randomness in the tunneling between the edge and parent superconductor can still reduce the gap compared to its ideal value; see Ref.~\onlinecite{Cole}.}, a comparatively large spectral gap\cite{PotterLee}, and the possibility of germinating exotic generalizations of Majorana zero modes known as `parafermions'\cite{Fendley,ZhangKane,parafermions}.

Encouraging experimental progress has recently transpired in both HgTe\cite{BernevigScienceHgTe,MolenkampHgTeSpinHallExp} and InAs/GaSb\cite{LiuQSH,Knez3} QSH-superconductor hybrids.\cite{Knez2,AmirSpinHallCurrent,LeoKouwenhovenSpinHall,KononovConductanceExp,AmirExpt2,MolenkampFJE2}  Strong superconducting proximity effects are now achievable in these materials.  Furthermore, unusual signatures in ac Josephson measurements\cite{MolenkampFJE2} (similar to Ref.~\onlinecite{FractionalJosephson}) have been interpreted as the `fractional Josephson effect'\cite{Kitaev-Unpaired-Majorana-1D-wire} that occurs uniquely in topological superconductors.  At this point it seems worthwhile to pursue complementary conductance probes reminiscent of those that have been widely utilized in related 1D platforms to detect Majorana-zero-mode signatures.\cite{Mourik25052012,das2012zeroBiasPeak,ZeroBiasPeakInSb,AaronZeroBiasPeak,Churchill,DeFranceschi,AliYazdaniMajoranaFermion,Exponential,BallisticMajorana}  To this end, the principal goals of this paper are to (1) identify universal transport fingerprints of topological superconductivity in QSH architectures, (2) propose relatively simple control experiments that provide sharp contrasts with trivial superconductivity, and (3) highlight the special role played by both interactions and time-reversal symmetry, which enrich the physics in interesting ways.

We apply renormalization-group techniques in two closely related setups to predict transport behavior at low energies, where the physics becomes largely insensitive to microscopic details.  The first setup, shown in Fig.~\ref{Device}, contains a \emph{grounded} superconductor that proximitizes an edge segment of length $L$.  We probe the paired region by sending charge through the adjacent gapless QSH edge states---modeled as Luttinger liquids that capture electron-electron interaction effects (see Ref.~\onlinecite{NonInteractingEdgeStateConductance} for a free-fermion treatment).  This setup is an edge counterpart to the nanowire experiments from Refs.~\onlinecite{Mourik25052012,das2012zeroBiasPeak,ZeroBiasPeakInSb,AaronZeroBiasPeak,Churchill,DeFranceschi,BallisticMajorana} that identify Majorana modes through zero-bias anomalies.  As in nanowires, when $L$ is much longer than the induced coherence length $\xi$ (so that the proximitized edge is in a meaningful sense topological) Andreev reflections dominate the low-energy transport leading to the familiar quantized zero-bias conductance\cite{Sengupta,ZeroBiasAnomaly3,ZeroBiasAnomaly4,ZeroBiasAnomaly6,Fidkowski2012,BeenakkerReview2,NonInteractingEdgeStateConductance}.  This conclusion persists over a broad range of interaction strength, including the experimentally relevant case of weak repulsion, and holds independent of whether time-reversal symmetry $\mathcal{T}$ is preserved or broken explicitly, e.g., by a weak magnetic field.  Time-reversal symmetry does, however, modify the allowed scattering processes and thus impacts both universal corrections to the quantized conductance and the critical interaction strength at which Andreev processes freeze out in favor of normal reflection. 

In nanowires, one can destroy the topological phase to contrast with the nontrivial behavior above simply by tweaking an external magnetic field or the electron density.  Similar methods can be adapted also to the QSH edge but are much less straightforward (a consequence of the `naturalness' of the topological phase in this setting).  For an alternative, more accessible control experiment we explore a proximitized edge that is simply too short to sustain topological superconductivity, i.e., $L \lesssim \xi$.  Here interactions and time-reversal symmetry yield more striking transport consequences.  Reference~\onlinecite{NonInteractingEdgeStateConductance} showed that zero-bias conductance in the free-fermion limit is non-universal.  In sharp contrast, arbitrarily weak repulsive interactions restore universality in a manner sensitively dependent on time-reversal symmetry: When $\mathcal{T}$ is present electrons at low energies perfectly transmit \emph{across} the impurity-like superconducting region---behavior generically absent in the analogous nanowire setup---whereas with broken $\mathcal{T}$ they perfectly backscatter at the interface.  These properties underlie nontrivial transport predictions, summarized in Fig.~\ref{ConductancePlots}, that clearly distinguish trivial and topological superconductivity.  

In the second setup that we explore, the edge is proximitized by a \emph{floating} superconductor, and magnetized regions on each end define an island with charging energy; see Fig.~\ref{DeviceWithChargingEnergy}.  Here we obtain an edge counterpart to recent nanowire experiments from Albrecht et al.\cite{Exponential} that reported unusual Coulomb blockade features---specifically $e$-periodic charging spectra---that originate from Majorana modes\cite{LiangFuTeleport}.  We show that when ferromagnetic barriers create the island in our QSH device, anomalous $e$-periodic charging also arises, but with an interesting twist: one can tune the offset charge to \emph{fractional} values by controlling the relative orientation of the ferromagnets (similar to the non-superconducting setup studied in Ref.~\onlinecite{qi_fractional_2008}).  Even more interestingly, when the barrier magnetizations arise from interaction-induced spontaneous $\mathcal{T}$ breaking, the charge-addition periodicity changes from $e$ to $e/2$.  The `fractional Coulomb blockade' in the latter case can be viewed as a manifestation of $\mathbb{Z}_4$ parafermion modes in the QSH edge\cite{ZhangKane,parafermions}.  By utilizing a mapping to earlier work by Kane and Fisher\cite{Kane-Fisher-PRB}, we also deduce the asymptotic boundary conditions imposed on the adjacent QSH edge by the island as a function of interaction strength.  

The remainder of the paper proceeds as follows.  Sections~\ref{GroundedSetup} through \ref{sec:conductance} develop the theory for the grounded-superconductor setup; in particular, Sec.~\ref{GroundedSetup} reviews the bosonized formulation that we use throughout, Secs.~\ref{sec:infinite_superconductor} and \ref{sec:superconductor} respectively analyze the `long' and `short' superconductor regimes, and Sec.~\ref{sec:conductance} extracts universal scaling forms for transport.  We then discuss the floating-superconductor case in Sec.~\ref{sec:FloatingSC} and finally conclude with a summary and outlook in Sec.~\ref{sec:conclusions}.

%%%%%%%%%%%%%%%%%%%%%%%%%%%%%%%%%%%%%%%%%%%%%%%%

\section{Grounded-superconductor setup}
\label{GroundedSetup}

\begin{figure}
	\includegraphics[width=8.5cm]{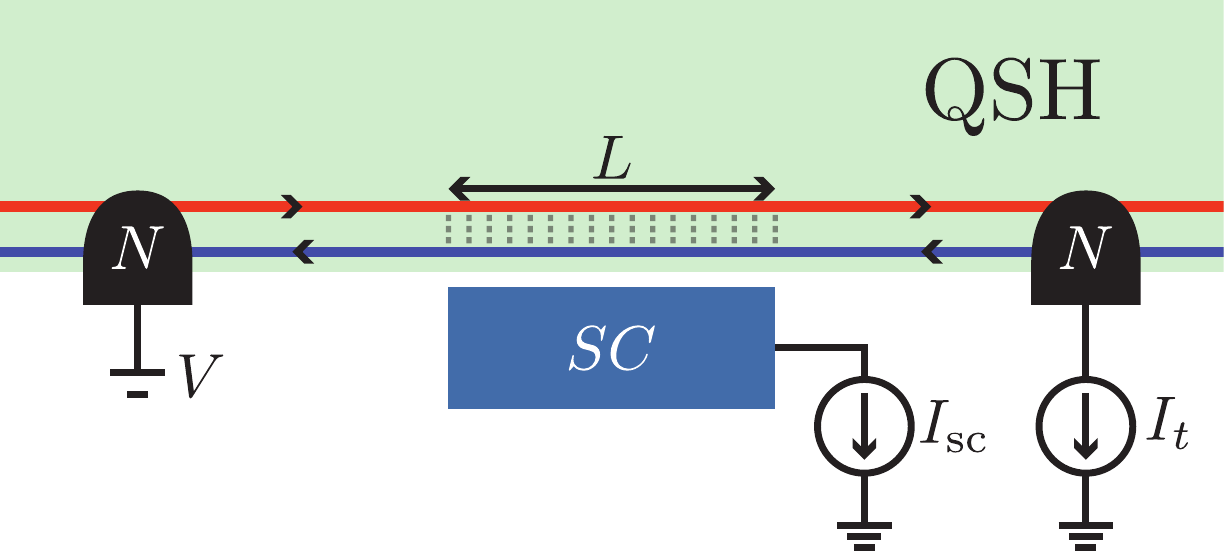}
	\caption{Quantum-spin-Hall setup proximitized by a ground superconductor that gaps out the helical edge states in a region of length $L$.  The induced superconductivity is probed by biasing one adjacent gapless edge state with a voltage $V$, and then measuring the currents $I_{\rm{SC}}$ flowing into the parent superconductor and $I_{\rm{t}}$ transmitted across it.  We predict universal forms for the corresponding conductances that depend on the interaction strength for the gapless edges, the size of the superconducting region relative to the induced coherence length $\xi$, and the presence or absence of time-reversal symmetry.
	}
	\label{Device}
\end{figure}

We start by discussing the gapless regions of the system that will serve as a lead for injecting charge into the grounded superconducting edge segment from Fig.~\ref{Device}.  Let $\psi_{R/L}(x)$ denote right/left-moving chiral fields describing the helical QSH edge modes at position $x$ along the boundary.  The edge Hamiltonian, including short-range interactions $H_{\rm int}$, can be written in these variables as
\begin{eqnarray}
  H_{\rm lead} &=& \int dx \left[\psi_{R}^{\dagger}(-i v \partial_x-\mu)\psi_{R}+\psi_{L}^{\dagger}(i v \partial_x-\mu)\psi_{L}\right]
  \nonumber \\
  &+& H_{\text{int}},
\end{eqnarray}
where $v$ is the non-interacting edge velocity and $\mu$ is the chemical potential.  Unless otherwise stated we assume $\mu \neq 0$ so that the Fermi momenta $\pm k_F$ are non-zero.  This property simplifies the structure of the system's effective low-energy theory and guarantees stability of the gapless modes even with strong interactions.\footnote{For example, a two-particle backscattering term $\sim \psi_R^\dagger \psi_R^\dagger \psi_L \psi_L$ (suitably regularized) can open a gap for strong repulsive interactions when $\mu = 0$, but is benign at $\mu \neq 0$.  Indeed from Eq.~\eqref{bosonize} one sees that in the former case such a term bosonizes to $\cos(4\theta)$ but oscillates spatially in the latter.}

To efficiently treat interactions in the `leads' we will exploit a bosonized representation, decomposing $\psi_{R/L}$ through
\begin{equation}
  \psi_{R/L} \sim e^{ \pm i k_F x} e^{i( \varphi \pm \theta)}\,,
  \label{bosonize}
\end{equation}
where the bosonic fields $\varphi$ and $\theta$ satisfy the commutation relation $ [\varphi(x), \theta(x^{\prime})] = i \pi \Theta(x - x^{\prime})$.  The edge electron density is then $\rho = \partial_x\theta/\pi$; for later use we note that commutation with $\varphi$ implies that $e^{im \varphi}$ increments the electric charge by $m$ units.  Bosonizing yields an effective low-energy Hamiltonian
\begin{equation}
H_{\rm  lead} =\frac{v}{2\pi} \int dx \left[ g\left(\partial_x \varphi \right)^2 +\frac{1}{g}\left(\partial_x \theta \right)^2 \right].
\label{LuttingerLiquid}
\end{equation}
The Luttinger parameter $g$ quantifies the interaction strength: $g = 1$ denotes the non-interacting limit while $g <1$ and $g>1$ respectively correspond to repulsive and attractive interactions.  We will be most interested in $g < 1$ since this regime is likely the most relevant for experiments.

Suppose now that a proximate superconductor generates a pairing gap at the edge between $x = 0$ and $x = L$; see Fig.~\ref{Device}.  We assume that the induced gap is larger than any other relevant energy scale in the problem (e.g., temperature $T$, bias voltage $V$, possible Zeeman energies, etc.).  In this case the superconducting region $(i)$ modifies the Hamiltonian for the adjacent gapless modes with generic symmetry-allowed terms and $(ii)$ in the asymptotic low-energy limit imposes certain boundary conditions on the leads that define boundary fixed points in renormalization group language.\cite{AffleckLL-SC}  The full effective low-energy Hamiltonian thus reads
\begin{equation}
H = H_{\rm lead} + H_B,
\end{equation}
where $H_B$ encodes perturbations involving fields at $x = 0$ and $x = L$.  In the limit $L/\xi \gg 1$ cross-couplings between opposite ends of the grounded superconductor can be neglected, and $H_B$ only includes local terms at each boundary separately.  However, with $L/\xi \lesssim 1$ additional terms that transfer charge \emph{across} the `short' superconductor are present.  As we will see, these terms qualitatively change the physics compared to the `long' superconductor case.

Apart from the size of the superconductor, time-reversal symmetry $\mathcal{T}$ also plays a central role throughout this paper.  Under $\mathcal{T}$ the fermions transform according to
\begin{equation}
  \mathcal{T}[\psi_R] = \psi_L,~~~~\mathcal{T}[\psi_L] = -\psi_R.
\end{equation}
Using Eq.~\eqref{bosonize}, and recalling antiunitarity, the bosonized fields in turn transform via
\begin{equation}
  \mathcal{T}[\varphi] = -\varphi - \pi/2,~~~~\mathcal{T}[\theta] = \theta+\pi/2.
  \label{eq:time_reversal}
\end{equation}
Perturbations in $H_B$, permissible boundary conditions, and their stability all depend sensitively on whether the system preserves $\mathcal{T}$.  In the following sections we will separately analyze the cases with and without time-reversal symmetry, both in the long- and short-superconductor limits.  We will specifically use renormalization group methods developed in Refs.~\onlinecite{KaneFisherTransportLuttingerliquid,Kane-Fisher-PRB,KaneFisherResonantTunneling,AffleckLL-SC,Fidkowski2012} to explore the stability of boundary conditions (i.e., fixed points) as a function of the Luttinger parameter $g$, and deduce the corresponding universal transport characteristics for the QSH/superconductor device.  Similar approaches have been used to study topological superconductivity in a variety of contexts\cite{Fidkowski2012,IanAffleckTSC,Komijani,MajoranaKramers2,KondoConductance,Zuo,KimLiu}.

%%%%%%%%%%%%%%%%%%%%%%%%%%%%%%%%%%%%%%%%%%%%%%%%

\section{Long superconductor limit $L \gg \xi$}
\label{sec:infinite_superconductor}

This section explores the case where the grounded superconducting region of the QSH edge is much longer than the induced coherence length.  The left and right interfaces then essentially decouple, so in this section we focus only on the left boundary at $x = 0$ for simplicity.\footnote{Tunneling across the superconductor remains small even when such processes are relevant in the renormalization-group sense, provided the flow is cut off by a small but finite temperature or voltage.} We note for later reference, however, that in the case of a floating superconductor charging effects can result in a coupling between the left and right interfaces via nontrivial domain-wall modes even in the limit $L\gg\xi$.  This scenario is discussed in Sec.~\ref{sec:FloatingSC}.

%%%%%%%%%%%%%%%%%%%%%%%%%%%%%%%%%%%%%%%%%%%%%%%%

\subsection{Fixed-point boundary actions}

The long-superconductor limit admits two natural types of boundary conditions in the asymptotic low-energy limit: An electron impinging on the superconductor from the adjacent gapless lead can, with unit probability, backscatter either as a hole (perfect Andreev reflection) or an electron (perfect normal reflection).

With perfect Andreev reflection the boundary condition for the fermionic fields takes the form $\psi_R(x = 0) = e^{i \alpha} \psi_L^{\dagger}(x = 0)$, which implies that the bosonized field $\varphi(x = 0)$ is pinned such that $e^{2i \varphi(x = 0)} = e^{i \alpha}$.  Note that time-reversal symmetry, when present, fixes $\alpha = \pm \pi/2$.  For assessing the stability of the fixed point defined by this boundary condition, it proves very useful to integrate out all fields away from $x = 0$ to obtain an effective theory for the remaining fluctuating field at the interface, $\Theta\equiv\theta(x = 0)$.\cite{Kane-Fisher-PRB,KaneFisherTransportLuttingerliquid}  This procedure yields the perfect-Andreev-reflection fixed-point action\cite{Fidkowski2012}
\begin{equation}
  S_{\rm A}[\Theta] = \int \frac{d \omega}{2\pi} \frac{|\omega|}{2\pi g} |\Theta_{\omega}|^2~~~~(\rm{Andreev~fixed~point}).
  \label{AndreevReflectionFixedPointAction}
\end{equation}

Since $\psi_R$ and $\psi_L$ form Kramers partners, perfect normal reflection necessitates either explicit $\mathcal{T}$ breaking, e.g., through an applied magnetic field, or spontaneous $\mathcal{T}$ breaking generated by interactions\cite{wu_helical_2006,xu_stability_2006}.  Perfect normal reflection imposes the boundary condition $\psi_R(x = 0) = e^{i\alpha^{\prime}} \psi_L(x = 0)$ for some arbitrary phase $\alpha'$.  It follows that the bosonized field $\theta(x = 0)$ is pinned, leaving $\Phi \equiv \varphi(x = 0)$ as the fluctuating variable at the interface. Once integrating out the gapless modes away from the interface, we similarly obtain the fixed-point action describing perfect normal reflection,
\begin{equation}
S_{\rm N}[\Phi] =  \int \frac{d \omega}{2\pi} \frac{g |\omega|}{2\pi} |\Phi_{\omega}|^2~~~~~(\rm{Normal~fixed~point}).
\label{NormalReflectionFixedPointAction}
\end{equation}

We next analyze the stability of these two fixed point actions in the case where time-reversal symmetry is present \emph{in the microscopic Hamiltonian} and then broken explicitly.  We will refer to the former as the `time-reversal-symmetric case', though we stress that perfect normal reflection can still occur in that regime via spontaneous $\mathcal{T}$ breaking.

%%%%%%%%%%%%%%%%%%%%%%%%%%%%%%%%%%%%%%%%%%%%%%%%

\subsection{Time-reversal-symmetric case}\label{Sec:TRIInfinitSC}

%%%%%%%%%%%%%%%%%%%%%%%%%%%%%%%%%%%%%%%%%%%%%%%%

\subsubsection{Stability of Andreev fixed point}
\label{AndreevStability1}

Suppose that the system begins at the perfect-Andreev-reflection fixed point described by Eq.~\eqref{AndreevReflectionFixedPointAction}.  When the Hamiltonian preserves time-reversal symmetry, the leading perturbation to the fixed-point action arises from two-particle backscattering generated at the superconductor interface:
\begin{eqnarray}
  \lambda_{{\rm 2bs}}(\psi_L^{\dagger} i\partial_x \psi_L^{\dagger}\psi_R i\partial_x \psi_R + \mathrm{H.c.}) \sim \lambda_{{\rm 2bs}}\cos{(4\Theta)}.
\end{eqnarray}
(Such a term is symmetry-allowed even for an edge chemical potential $\mu\neq 0$ due to broken translation invariance at the boundary.)
The coupling $\lambda_{{\rm 2bs}}$ flows under renormalization according to
\begin{equation}
  \frac{d \lambda_{{\rm 2bs}}}{d l} = \left(1-8g \right) \lambda_{{\rm 2bs}},
\label{ConductanceTRIAR}
\end{equation}
with $l$ a logarithmic rescaling factor.  In the parenthesis, the factor of unity appears because our boundary problem corresponds to zero spatial dimensions and one imaginary time dimension, while $8g$ is the scaling dimension of the $\cos(4\Theta)$ perturbation.  We thus see that two-particle backscattering destabilizes the Andreev fixed point only for very strong repulsive interactions with $g < 1/8$.  Since this perturbation promotes normal reflection and favors pinning $\Theta$, it is natural to anticipate that for any $g < 1/8$ the system flows to the perfect-normal-reflection fixed point---thereby breaking time-reversal symmetry spontaneously.

%%%%%%%%%%%%%%%%%%%%%%%%%%%%%%%%%%%%%%%%%%%%%%%%

\subsubsection{Stability of normal fixed point}
\label{normalTinvariant}

Imagine now that, due to spontaneous $\mathcal{T}$ breaking, the system instead begins at the normal-reflection fixed point described by Eq.~\eqref{NormalReflectionFixedPointAction}.  For consistency with our results above, we expect that this fixed point is stable for $g < 1/8$, and further that for any $g>1/8$ a physical perturbation drives a flow back to the Andreev fixed point.  Because $\Theta$ is pinned due to normal-reflection boundary conditions, perturbations at the normal fixed point should take the bosonized form $\tilde{\lambda}_k \cos(k\Phi + \delta_k)$ for some phases $\delta_k$.  These couplings flow according to
\begin{equation}
\frac{d \tilde{\lambda}_k}{d l} =  \left(1-\frac{k^2}{2g} \right)\tilde{\lambda}_k
\label{lambdak}
\end{equation}
and are relevant for $g>k^2/2$.  What, then, are the physical values of $k$?

A naive guess for the leading perturbation to the action is a local pairing term generated at the superconductor interface,
\begin{eqnarray}
  \psi_R \psi_L + \mathrm{H.c.} \sim \sin{(2\Phi)},
  \label{pairing}
\end{eqnarray}
which corresponds to $k = 2$ above.  This process indeed promotes Andreev reflection but becomes relevant only for strong attractive interactions with $g>2$.

Curiously, a perturbation with $k = 1/2$, which we hereafter denote by
\begin{equation}
  \lambda_{\rm pf}\cos\left(\frac{\Phi}{2}-\delta\right),
  \label{lambdapf}
\end{equation}
is necessary to destabilize the normal fixed point for any $g > 1/8$.  A duality transformation hints that this is indeed the `correct' perturbation to exploit here, since the $\lambda_{\rm pf}$ term at the normal-reflection fixed point is dual to the two-particle backscattering term $\cos(4\Theta)$ at the Andreev fixed point. The analysis closely follows Ref.~\onlinecite{Fidkowski2012}, so we relegate details to Appendix \ref{APP:duality}.  On the other hand, no local combination of fermion fields $\psi_{R/L}$ acting at the interface generates Eq.~\eqref{lambdapf}, so at first glance this term seems unphysical (it changes the charge in the gapless edge by $\pm e/2$). Interestingly, such fractional charge transfers can nevertheless become possible in strongly interacting QSH edge systems\cite{maciejko_kondo_2009}. The physical picture here is that the $\lambda_{\rm pf}$ term arises not simply from ordinary electron degrees of freedom, but rather from hybridization between the lead and an additional `parafermion zero mode' that appears dynamically at the interface when interactions are strong.

\begin{figure}
\includegraphics[width=8.5cm]{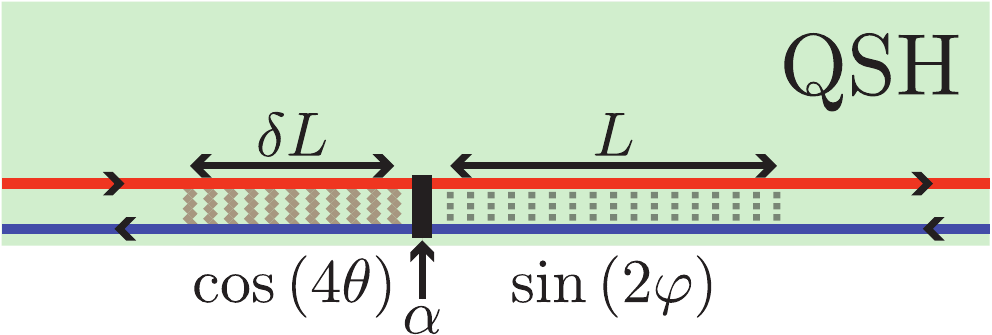}
\caption{Variation of Fig.~\ref{Device} that generates perfect-normal-reflection boundary conditions without explicitly breaking time-reversal symmetry.  The gapless edge on the left side and the superconductor are now bridged by a region of length $\delta L$ in which two-particle backscattering [i.e., $\cos(4\theta)$] violates time-reversal \emph{spontaneously}---thus naturally allowing perfect normal reflection.  The domain wall separating the magnetic and superconducting regions binds a $\mathbb{Z}_4$ parafermion zero mode $\alpha$.\cite{ZhangKane,parafermions}  As $\delta L$ shrinks, hybridization between the gapless edge and the zero mode allows resonant transfer of $e/2$ charges [see Eq.~\eqref{lambdapf}]; for the long-superconductor case with $g>1/8$, such a perturbation destabilizes the perfect-normal-reflection fixed point and restores perfect Andreev reflection at low energies.
}
\label{Parafermion}
\end{figure}

To see this, suppose that we access the normal fixed point using the modified (but isosymmetric) edge geometry of Fig.~\ref{Parafermion}.  Here an extended region of length $\delta L$ with a relevant $\cos(4\theta)$ two-particle backscattering term bridges the superconductor and the gapless edge states.  (Relevance requires that the chemical potential $\mu$ for the $\delta L$ segment vanishes, though elsewhere we still assume $\mu \neq 0$.)  It is useful to define a local magnetization order parameter $M \equiv \psi_L^\dagger \psi_R +H.c. \sim \cos{(2\theta)}$ that is odd under time reversal [recall Eq.~\eqref{eq:time_reversal}].  Upon pinning of $\theta$ by $\cos(4\theta)$, the intervening region takes on one of two non-zero values for $\langle M\rangle$---breaking $\mathcal{T}$ spontaneously as we assumed above.  Since the two magnetizations yield identical energies, the gapped domains collectively host a larger ground-state degeneracy compared to the usual ferromagnet-superconductor configurations studied earlier by Fu and Kane\cite{Fu-Kane-QSH2}.  In the latter case Majorana zero modes---which can absorb electrons with no energy cost---bind to domain walls between pairing and magnetically gapped regions.  The additional degeneracy in our setup promotes the Majoranas to more exotic $\mathbb{Z}_4$ parafermion zero modes that can similarly absorb fractional $e/2$ charges without energy penalty\cite{ZhangKane,parafermions,ParafermionReview}.

\begin{figure}
\includegraphics[width=8.5cm]{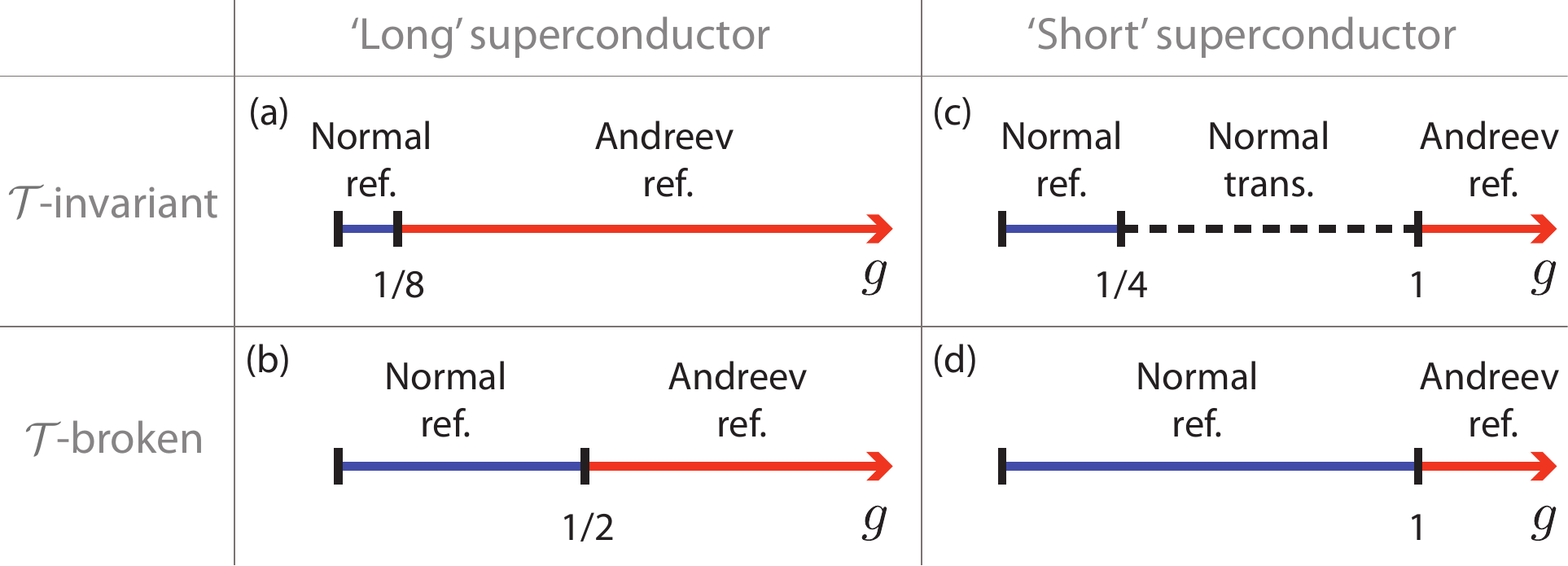}
\caption{Phase diagrams of the grounded superconductor/helical edge setup as a function of the Luttinger parameter $g$.  Even for weak repulsive interactions expected to be relevant for experiment ($g$ slightly smaller than one), three stable boundary fixed points are accessible depending on the size of the superconductor and whether time-reversal $\mathcal{T}$ is preserved or broken explicitly.  Each fixed point yields distinct, universal values for the conductance $G_t$ across the superconductor and $G_{SC}$ through the superconductor in the zero-bias, zero-temperature limit.  Specifically, the Andreev reflection, normal reflection, and normal transmission fixed points respectively give $(G_t,G_{SC}) = (0,2e^2/h), (0,0)$, and $(e^2/h,0)$.  }
\label{PhaseDiagram}
\end{figure}

As the width $\delta L$ shrinks towards zero, coupling to the parafermion zero mode allows charge $e/2$ excitations to resonantly tunnel between the gapless edge and the adjacent domain wall.  As derived in Appendix \ref{APP:Parafermion}, such processes generate precisely the term in Eq.~\eqref{lambdapf} that destabilizes the normal-reflection fixed point for $g>1/8$, driving a flow back to the Andreev fixed point where $\Phi$ is instead fixed.  [The derivation in the appendix shows that the shift $\delta$ in Eq.~\eqref{lambdapf} is actually an operator that transforms nontrivially under $\mathcal{T}$, though the scaling dimension for $\lambda_{\rm pf}$ is unaffected.  This is actually essential for obtaining a $\mathcal{T}$-invariant term as is evident from Eq.~\eqref{eq:time_reversal}.]  Thus our analyses of the normal and Andreev fixed points cleanly gel with one another, leading to the phase diagram shown in Fig.~\ref{PhaseDiagram}(a).

%%%%%%%%%%%%%%%%%%%%%%%%%%%%%%%%%%%%%%%%%%%%%%%%

\subsection{Explicit time-reversal-broken case}

When time-reversal symmetry is broken \emph{explicitly} (e.g., by a weak external magnetic field), our edge problem maps precisely onto a strictly one-dimensional Luttinger liquid/topological superconductor junction explored in Ref.~\onlinecite{Fidkowski2012}.  We will, however, briefly highlight the main results to emphasize the new features of the $\mathcal{T}$-invariant case discussed in the previous subsection.

%%%%%%%%%%%%%%%%%%%%%%%%%%%%%%%%%%%%%%%%%%%%%%%%

\subsubsection{Stability of Andreev fixed point}

With explicit $\mathcal{T}$ breaking, the leading perturbation to the Andreev fixed point corresponds to ordinary single-particle backscattering at the interface with the superconductor:
\begin{equation}
  \lambda_{{\rm 1bs}}(\psi_L^{\dagger}\psi_R+H.c.)\sim \lambda_{{\rm 1bs}}\cos(2\Theta).
\end{equation}
The coupling constant $\lambda_{{\rm 1bs}}$ renormalizes according to
\begin{equation}
  \frac{d\lambda_{{\rm 1bs}}}{dl}=(1-2g)\lambda_{{\rm 1bs}}
  \label{lambda1bs_flow}
\end{equation}
and thus generates a flow to the normal fixed point for $g < 1/2$.  Compared to the $\mathcal{T}$-invariant case, the Andreev fixed point is stable over a more restricted range of interactions simply because lower-order backscattering processes are now available.

%%%%%%%%%%%%%%%%%%%%%%%%%%%%%%%%%%%%%%%%%%%%%%%%

\subsubsection{Stability of normal fixed point}

For consistency we expect a stable perfect-normal-reflection fixed point now only at $g < 1/2$.  Perturbations to this fixed point once again take the form $\tilde{\lambda}_k \cos(k\Phi+\delta_k)$, with flow equations given in Eq.~\eqref{lambdak}.  In this case the $k = 1$ term---which we will denote $\lambda_{\rm M}\cos(\Phi+\delta')$---represents the physical process that becomes relevant and drives a flow back to the Andreev fixed point for $g>1/2$.

Notice, however, that the $\lambda_{\rm M}$ perturbation changes the electron number in the gapless edge by $\pm 1$, whereas one might naively expect the adjacent gapped superconductor to absorb only Cooper pairs at low energies.  One can see why this term is indeed physically admissible by accessing the normal fixed point using the geometry of Fig.~\ref{Parafermion} where a region gapped by $\cos(4\theta)$ separates the gapless and superconducting parts of the edge.  Because time-reversal symmetry is now broken explicitly, a $\cos(2\theta)$ term will generically appear as well---lifting the degeneracy between the two magnetization values that would otherwise be chosen spontaneously in the intervening edge segment.  This degeneracy lifting correspondingly demotes the $\mathbb{Z}_4$ parafermion zero mode trapped at the domain wall to a Majorana zero mode as in the usual Fu-Kane setup\cite{Fu-Kane-QSH2}.  Hybridization with the Majorana zero mode allows the gapless edge to coherently transfer single electrons to the domain wall, thereby generating the $\lambda_{\rm M}$ term invoked above.  Thus here too the Andreev and normal fixed point analyses agree, yielding the phase diagram in Fig.~\ref{PhaseDiagram}(b).

Table \ref{infiniteSCPerturbation} summarizes the long-superconductor results for the $\mathcal{T}$-invariant and $\mathcal{T}$-broken cases.  We see that the main effect of time-reversal symmetry is to forbid elastic single-electron backscattering at the interface, thus extending the stability window of the Andreev fixed point down to much stronger repulsive interaction strengths.  Next we turn to the short-superconductor limit, where time-reversal symmetry plays a more prominent role even for a weakly interacting edge.

\begin{table}
\begin{tabular}{|l|l|l|l|}
\hline
symmetry &action& perturbation & dim \\
\hline
\multirow{2}{*}{$\mathcal{T}$-invariant} &$S_A$ & two-particle backscattering & $8g$\\
\cline{2-4}
&$S_N$ &$\mathbb{Z}_4$ parafermion hybridization & $1/(8g)$\\ \hline
\multirow{2}{*}{$\mathcal{T}$-broken} &$S_{A}$  & one-electron backscattering& $2g$\\
\cline{2-4}
&$S_N$ & Majorana hybridization & $1/(2g)$\\
\hline
\end{tabular}\\
\caption{Summary of fixed-point actions and their leading perturbations in the `long' superconductor limit.  We denote the actions describing perfect-normal-reflection and perfect-Andreev-reflection boundary conditions as  $S_N$ and $S_A$, respectively. Each fixed point is stable when the scaling dimensions (right column) of the corresponding perturbations are larger than one.
}
\label{infiniteSCPerturbation}
\end{table}

%%%%%%%%%%%%%%%%%%%%%%%%%%%%%%%%%%%%%%%%%%%%%%%%

\section{Short superconductor limit $L\lesssim \xi$}
\label{sec:superconductor}

%%%%%%%%%%%%%%%%%%%%%%%%%%%%%%%%%%%%%%%%%%%%%%%%

\subsection{Fixed-point boundary actions}

When the length $L$ of the pairing-gapped edge segment in Fig.~\ref{Device} is comparable to the induced coherence length, the superconductor mimics a quantum impurity that can mediate transport between the adjacent gapless `leads' on either side.  Thus here it is essential to keep track of both interfaces simultaneously.  We proceed as above, identifying boundary fixed-point actions and then exploring their stability to physical perturbations.  Let $\Phi_1, \Theta_1$ denote the bosonized fields $\varphi,\theta$ at the left superconductor interface, with $\Phi_2,\Theta_2$ denoting the boundary fields at the right interface.  We will specifically study the stability of three natural types of boundary conditions:

$(i)$ Perfect Andreev reflection at each interface separately, described by the boundary action
\begin{equation}
  S_{A\oplus A} = S_{A}[\Theta_1] +S_{A}[\Theta_2].
\end{equation}

$(ii)$ Perfect normal reflection at each interface separately, where the superconductor effectively `cuts' the edge at low energies.  This fixed point is described by
\begin{equation}
  S_{N\oplus N} = S_{N}[\Phi_1] +S_{N}[\Phi_2].
\end{equation}
These first two cases straightforwardly generalize the fixed-point theories defined in Eqs.~\eqref{AndreevReflectionFixedPointAction} and \eqref{NormalReflectionFixedPointAction}.

$(iii)$ Perfect normal \emph{transmission}, wherein incident electrons tunnel past the superconductor with unit probability.  Here the superconducting `impurity' becomes invisible at low energies, so that the boundary fields match on both sides: $\Phi_1=\Phi_2 \equiv \Phi$ and $\Theta_1=\Theta_2\equiv \Theta$.\footnote{Technically, time-reversal symmetry allows for a more general boundary condition with $\Theta_1 = \Theta_2 + \alpha$ for arbitrary real $\alpha$.  We simply set $\alpha = 0$ since this parameter does not play a role in our analysis.}  Integrating out fields away from the boundary yields the normal transmission fixed-point action
\begin{equation}
  S_T[\Phi,\Theta] = \int \frac{d \omega}{2 \pi} \frac{|\omega|}{\pi} \left( g |\Phi_\omega|^2 + g^{-1} |\Theta_\omega|^2 \right).
  \label{TransmissionFixedPointAction}
\end{equation}

Other boundary conditions are also possible, most notably perfect \emph{crossed} Andreev reflection, wherein an incident electron from one end of the superconductor transmits with unit probability as a hole in the other.  Reference~\onlinecite{MajoranaKramers2} showed that this boundary condition can be stable in a related interacting system that supports a Kramers pair of Majorana zero modes\footnote{For an interesting discussion of the non-interacting limit of that setup see Ref.~\onlinecite{MajoranaKramers1}}; a chiral analogue can also appear when quantum Hall edge states serve as a lead\cite{Interferometry1,Interferometry2,ClarkeCircuits}.  As discussed in Appendix~\ref{CAR}, however, our setup is unlikely to realize such a fixed point in practice; thus for the remainder of this section we focus only on cases $(i)$-$(iii)$ above.  

%%%%%%%%%%%%%%%%%%%%%%%%%%%%%%%%%%%%%%%%%%%%%%%%

\subsection{Time-reversal-symmetric case}\label{Sec:TRIshortSC}

%%%%%%%%%%%%%%%%%%%%%%%%%%%%%%%%%%%%%%%%%%%%%%%%

\subsubsection{Stability of Andreev$\oplus$Andreev fixed point}

Consider first the fixed point characterizing perfect Andreev reflection at each interface.  With time-reversal present in the microscopic Hamiltonian, the lowest-order allowed perturbation acting at a given boundary [$\cos(4\Theta_{1,2})$] arises from two-particle backscattering---which we saw earlier requires very strong repulsive interactions to become relevant.  The short-superconductor limit, however, additionally permits a $\mathcal{T}$-invariant term that couples the two boundaries,
\begin{equation}
  \lambda_t \cos(\Theta_1-\Theta_2 + \chi)
  \label{lambdat}
\end{equation}
for some non-universal phase $\chi$, and is more effective at destabilizing the Andreev boundary conditions.  Microscopically, tunneling of electrons across the superconductor (e.g., $\psi_{R1}^{\dagger}\psi_{R2} + \psi_{L1}^{\dagger}\psi_{L2}+H.c.$) generates precisely such a term.\footnote{This identification is not unique, since upon using the Andreev boundary condition $\psi_{R} \sim \psi_L^\dagger$ other microscopic processes can also give rise to the same bosonized perturbation $\propto \cos(\Theta_1-\Theta_2+\chi)$.}   Under renormalization, we have
\begin{equation}
  \frac{d\lambda_t}{dl} = (1 - g)\lambda_t,
\label{eq:normal_transmission_scaling}
\end{equation}
indicating that with arbitrarily weak repulsive interactions (i.e., at any $g < 1$) $\lambda_t$ is relevant and destabilizes the independent Andreev boundary conditions.  Since perfect normal transmission is a natural candidate fixed point for the system to then flow towards, we next turn to the stability of that boundary condition.

%%%%%%%%%%%%%%%%%%%%%%%%%%%%%%%%%%%%%%%%%%%%%%%%

\subsubsection{Stability of normal transmission fixed point}
\label{NormalTransmissionStability}

Starting from the perfect normal transmission fixed point, one can always add a boundary perturbation $\lambda_{A}\sin(2\Phi)$.  This term encodes Andreev-reflection processes at either superconductor boundary; cf.~Eq.~\eqref{pairing}.  The coupling flows via
\begin{equation}
  \frac{d\lambda_{A}}{dl} =  \left( 1 - \frac{1}{g} \right) \lambda_{A}.
\end{equation}
[Notice that the flow equation differs from that quoted earlier for Eq.~\eqref{pairing} because we are now using a different fixed-point action.  Compare Eqs.~\eqref{NormalReflectionFixedPointAction} and \eqref{TransmissionFixedPointAction}.]  With attractive interactions $g>1$, $\lambda_A$ is relevant and naturally drives a flow back to the Andreev$\oplus$Andreev fixed point, which we saw above is stable in that parameter regime.

The normal transmission fixed point is also unstable for strong repulsive interactions.  Two-particle backscattering, $\lambda_{{\rm 2bs}}\cos(4\Theta)$, now flows according to
\begin{equation}
  \frac{d\lambda_{{\rm 2bs}}}{dl} =\left( 1 - 4g \right)\lambda_{{\rm 2bs}}.
\end{equation}
For $g<1/4$ this coupling is relevant and breaks $\mathcal{T}$ spontaneously.  The edge is then effectively sliced in two, and the system flows to the normal$\oplus$normal fixed point.

%%%%%%%%%%%%%%%%%%%%%%%%%%%%%%%%%%%%%%%%%%%%%%%%

\subsubsection{Stability of normal$\oplus$normal fixed point}
\label{sec:normal_normal_T}

Consider next a straightforward generalization of Fig.~\ref{Parafermion} in which we add a narrow region with a relevant $\cos(4\theta)$ term to each end of the short superconductor.  Time-reversal is then broken spontaneously at the two boundaries, allowing us to enter the normal$\oplus$normal fixed point without explicitly violating $\mathcal{T}$.  The most relevant term with which we can perturb this fixed point is
\begin{equation}
  \lambda_{e/2} \cos\left(\frac{\Phi_1-\Phi_2}{2}\right).
  \label{NNperturbation}
\end{equation}
This term $(i)$ transfers charge $e/2$ across the superconductor, $(ii)$ is gauge invariant and can be written in terms of currents [i.e., $e^{i (\Phi_2-\Phi_1)/2} = e^{i \int_x \partial_x \varphi}$], and $(iii)$ reverses the spontaneously chosen magnetization in the $\cos(4\theta)$ regions\footnote{The magnetizations in the left and right regions are \emph{not} independent since we are working with a short superconductor.  This point closely relates to the absence of $\mathbb{Z}_4$ parafermion zero modes discussed below.} and hence keeps the system within the low-energy subspace of interest.  Equation~\eqref{NNperturbation} thus constitutes a physically admissible perturbation that favors `resewing' the edge back together.  Under renormalization we have
\begin{equation}
  \frac{d\lambda_{e/2}}{dl} =\left( 1 - \frac{1}{4g} \right)\lambda_{e/2}.
\end{equation}
Consequently, for $g>1/4$ $\lambda_{e/2}$ is relevant and generates a flow back to the perfect normal transmission fixed point.

Note that in Sec.~\ref{normalTinvariant} we saw that hybridization between the gapless edge and a $\mathbb{Z}_4$ parafermion zero mode destabilized perfect-normal-reflection boundary conditions at $g>1/8$.  One might initially expect such a perturbation to be operative also at the normal$\oplus$normal fixed point studied here, but that is not so: zero-modes have no integrity in the short-superconductor limit since they will generically couple with their partner at the neighboring domain wall.  Thus Eq.~\eqref{NNperturbation} indeed represents the leading perturbation available.

Once again our stability analyses for the different fixed points are perfectly consistent as summarized in the upper rows of Table~\ref{shortSCPerturbationTRI}; Fig.~\ref{PhaseDiagram}(c) depicts the resulting phase diagram.

\begin{table}
	\begin{tabular}{|l|l|l|l| }
		\hline
		symmetry &action& perturbation & dim \\
		\hline
		\multirow{4}{*}{$\mathcal{T}$-invariant} &$S_{A \oplus A}$  & electron tunneling & $g$\\		
		\cline{2-4}
		&$S_T$ & Andreev reflection & $1/g $\\
		& & two-particle backscattering & $4g$\\
		\cline{2-4}
		&$S_{N \oplus N}$  & $e/2$ tunneling & $1/(4g)$\\
		\hline
		\multirow{4}{*}{$\mathcal{T}$-broken} &$S_{A \oplus A}$  & electron tunneling & $g$\\		
		\cline{2-4}
		&$S_T$ & Andreev reflection & $1/g $\\
		& & one-electron backscattering & $g$\\
		\cline{2-4}
		&$S_{N \oplus N}$  & electron tunneling & $1/g$\\
		\hline
	\end{tabular}\\
	\caption{Summary of fixed-point actions and their leading perturbations in the `short' superconductor limit.  Here $S_{A \oplus A}$ and $S_{N \oplus N}$ respectively describe perfect normal reflection and perfect Andreev reflection at each interface separately, while $S_T$ describes perfect normal transmission across the superconductor.  Stability requires that the scaling dimensions (right column) of the leading perturbations exceed one.  Notice that with explicitly broken time-reversal symmetry, the stability window for perfect normal transmission disappears completely.  }
	\label{shortSCPerturbationTRI}
\end{table}

%%%%%%%%%%%%%%%%%%%%%%%%%%%%%%%%%%%%%%%%%%%%%%%%

\subsection{Explicit time-reversal-broken case}

%%%%%%%%%%%%%%%%%%%%%%%%%%%%%%%%%%%%%%%%%%%%%%%%

\subsubsection{Stability of Andreev$\oplus$Andreev fixed point}

Breaking time-reversal symmetry explicitly does not alter the Andreev$\oplus$Andreev fixed point's stability window.  There are no physical perturbations with scaling dimension smaller than that of single-electron tunneling across the superconductor---regardless of the presence of $\mathcal{T}$---and so this fixed point continues to be stabilized by attractive interactions ($g>1$) and destabilized by repulsion ($g<1$).

%%%%%%%%%%%%%%%%%%%%%%%%%%%%%%%%%%%%%%%%%%%%%%%%

\subsubsection{Stability of normal transmission fixed point}

Stability of the normal transmission fixed point, by contrast, is obliterated by explicit $\mathcal{T}$ breaking.  We previously saw that local Andreev reflection processes destabilize the fixed point for $g>1$.  But now single-electron backscattering at each superconductor interface is also permitted.  In bosonized form the perturbation reads $\lambda_{{\rm 1bs}} \cos(2\Theta)$, which carries scaling dimension $g$ and becomes relevant for any $g<1$.  Thus the extended window of perfect normal transmission arising in the $\mathcal{T}$-invariant case indeed disappears.

%%%%%%%%%%%%%%%%%%%%%%%%%%%%%%%%%%%%%%%%%%%%%%%%

\subsubsection{Stability of normal$\oplus$normal fixed point}
\label{sec:normal_normal_broken}

The demise of stable normal transmission boundary conditions is accompanied by an enhanced stability window for the normal$\oplus$normal fixed point.  Let us again access the latter fixed point by magnetizing a portion of the edge on both sides of the superconductor (recall Fig.~\ref{Parafermion}).  Crucially, with explicit $\mathcal{T}$ breaking the magnetization is no longer spontaneously chosen.  Equation~\eqref{NNperturbation} thus does not constitute a legitimate low-energy perturbation; that operator flips the magnetization from its preferred orientation and places the system into a high-energy configuration.  In this case the leading perturbations are instead ordinary electron tunneling $\lambda_t \cos(\Phi_1-\Phi_2)$ and crossed Andreev reflection $\lambda_{\rm CAR}\cos(\Phi_1+\Phi_2)$.  Both terms become relevant only at $g>1$, so that the normal$\oplus$normal fixed point is now stable for any $g<1$.

Our results for the setup with explicit $\mathcal{T}$ breaking are summarized in Table~\ref{shortSCPerturbationTRI}.  Figure~\ref{PhaseDiagram}(d) shows the corresponding phase diagram---which departs dramatically from the $\mathcal{T}$-invariant case in Fig.~\ref{PhaseDiagram}(c).  For perspective on these findings, note that the broken-$\mathcal{T}$, short-superconductor setup closely resembles the single-channel Luttinger liquid with a point impurity studied in classic work by Kane and Fisher\cite{KaneFisherTransportLuttingerliquid}.  In the latter problem the (non-superconducting) impurity cuts the Luttinger liquid in two at low energies for $g<1$ (just as in our problem) but renormalizes to zero for $g>1$---generating perfect normal transmission.  When the impurity superconducts, however, attractive interactions instead give way to perfect Andreev reflection over the entire interval $g>1$ as found above.

%%%%%%%%%%%%%%%%%%%%%%%%%%%%%%%%%%%%%%%%%%%%%%%%

\section{Universal conductance properties}
\label{sec:conductance}

Having mapped out the phase diagrams, we are now in position to extract transport predictions for our quantum-spin-Hall system proximitized by a grounded superconductor.  We are specifically interested in the low-temperature voltage dependences of the conductances $G_{\rm SC} = I_{\rm SC}/V$ and $G_t = I_t/V$; here $V$ is the bias voltage while $I_{SC}$ and $I_t$ respectively denote the currents collected through the superconductor, and from the gapless edge just past it (see Fig.~\ref{Device}).  When analyzing these quantities we will assume that conduction between the two normal leads in Fig.~\ref{Device} arises predominantly across the superconductor, and \emph{not} through the complementary ungapped part of the QSH edge; i.e., electrons do not take the `long way around'.  This assumption is justified provided the latter paths are much longer than the inelastic scattering length---which should not be difficult to satisfy in practice.   

We will also restrict our attention to edges with weak repulsive interactions, i.e., $g$ slightly smaller than 1, since this regime is expected to be most experimentally relevant.  The phase diagrams in Fig.~\ref{PhaseDiagram} show that three types of boundary conditions can be stable, depending on the size of the superconducting region and whether time-reversal is intact: perfect Andreev reflection, perfect normal reflection, and perfect normal transmission.  In the extreme limit $V\rightarrow 0$ (and with temperature $T \rightarrow 0$) one can immediately deduce the conductances simply from the boundary conditions imposed asymptotically.  For the long-superconductor case, we have
\begin{equation}
  G_{\rm SC} = \frac{2e^2}{h},~~G_{t} = 0,~~({\rm long~SC}, V\rightarrow 0),
  \label{Glong}
\end{equation}
independent of the presence or absence of $\mathcal{T}$ symmetry.  The factor of 2 in $G_{\rm SC}$ appears because each incident electron from the gapless edge Andreev reflects and injects a Cooper pair into the superconductor with unit probability. \footnote{The presence of Fermi-liquid leads that inject charge into the edge states yields a conductance of $2e^2/h$ rather than the intrinsic value for a Luttinger liquid, for which an additional factor of $g$ appears (see Refs.~\onlinecite{MichaelStoneConductance,Safi1995})}  For a short superconductor with time-reversal symmetry, stable perfect normal transmission instead yields
\begin{equation}
  G_{\rm SC} = 0,~~G_{t} = \frac{e^2}{h},~~(\mathcal{T}{\text-}{\rm invariant~short~SC}, V\rightarrow 0).
  \label{Gshort_T}
\end{equation}
And finally, for a short superconductor with explicitly broken $\mathcal{T}$, perfect normal reflection sets in so that
\begin{equation}
  G_{\rm SC} = 0,~~G_{t} = 0,~~(\mathcal{T}{\text-}{\rm broken~short~SC}, V\rightarrow 0).
  \label{Gshort_noT}
\end{equation}

In what follows we will obtain corrections to Eqs.~\eqref{Glong} through \eqref{Gshort_noT} at low (and sometimes intermediate) bias voltages to predict precisely how the conductances approach these fixed-point values as $V$ decreases towards zero.  These corrections arise predominantly from the leading perturbation at the respective fixed points, and may be calculated using Keldysh formalism (for recent applications in a related context see Refs.~\onlinecite{RomanLutchynKeldyshApproach,Zazunov}).  We will alternatively use scaling arguments to deduce \emph{universal} power-law corrections in each case, as done, e.g., in Refs.~\onlinecite{KaneFisherTransportLuttingerliquid,Kane-Fisher-PRB,KaneFisherResonantTunneling,AffleckLL-SC,Fidkowski2012}.  

%%%%%%%%%%%%%%%%%%%%%%%%%%%%%%%%%%%%%%%%%%%%%%%%

\subsection{Long-superconductor limit}

In the long-superconductor limit, $G_t$ remains zero to an excellent approximation over an extended range of bias voltages.  Thus we simply focus on finite-voltage corrections to $G_{\rm SC}$.  We saw in Sec.~\ref{sec:infinite_superconductor} that the leading perturbation to the Andreev fixed point that encodes normal reflection---thereby suppressing $G_{\rm SC}$---is two-particle backscattering
\begin{equation}
  \delta H_{\mathcal{T}} = \lambda_{{\rm 2bs}}\cos{(4\Theta)}
\end{equation}
when time-reversal is present and single-electron backscattering
\begin{equation}
  \delta H_{{\rm no}\text{-}\mathcal{T}} =  \lambda_{{\rm 1bs}}\cos(2\Theta).
\end{equation}
otherwise.  The flow equations in Eqs.~\eqref{ConductanceTRIAR} and \eqref{lambda1bs_flow} determine the renormalized couplings $\lambda_a$ (with $a = {\rm 2bs}$ or $\rm{1bs}$) at an energy scale $E$.  Writing the logarithmic rescaling factor as $l = \ln(\Lambda/E)$, with $\Lambda$ a cutoff of order the induced pairing gap for the superconducting region, one finds
\begin{equation}
  \lambda_a(E) = \lambda^{(0)}_a \left(E/\Lambda\right)^{\Delta_a-1}.
  \label{lambdaE}
\end{equation}
Here $\lambda^{(0)}_a$ is the bare value of the coupling while $\Delta_a$ is the scaling dimension for the corresponding operator.

The leading correction $\delta G_{\rm SC}$ to the conductance coming from the above perturbations arises at second order in the couplings.  Setting the energy scale $E$ equal to the bias voltage $V$ at which we probe the system therefore yields $\delta G_{\rm SC} \propto V^{2(\Delta_a-1)}$ and hence scaling forms\footnote{Exactly the same logic allows one to deduce temperature dependence at zero bias voltage instead of voltage dependence at zero tempeature; for the former case one simply swaps $V\rightarrow T$ in the power-laws obtained here and below.}
\begin{equation}
  G_{\rm SC}(V) \sim \begin{cases}
               \frac{2e^2}{h}\left[1-(V/V_{\rm 2bs})^{2(8g-1)}\right],~\mathcal{T}\text{-invariant} \\
               \frac{2e^2}{h}\left[1-(V/V_{\rm 1bs})^{2(2g-1)}\right],~\mathcal{T}\text{-broken}
            \end{cases}.
   \label{GSClong}
\end{equation}
On the right side $V_{\rm 2bs}$ and $V_{\rm 1bs}$ are non-universal voltage scales determined by the bare amplitude for the associated backscattering processes.  The conductance corrections in Eq.~\eqref{GSClong} apply in the small-voltage regime, i.e., $V \ll V_{\rm 2bs/1bs}$.

\begin{figure*}
\includegraphics[width=17cm]{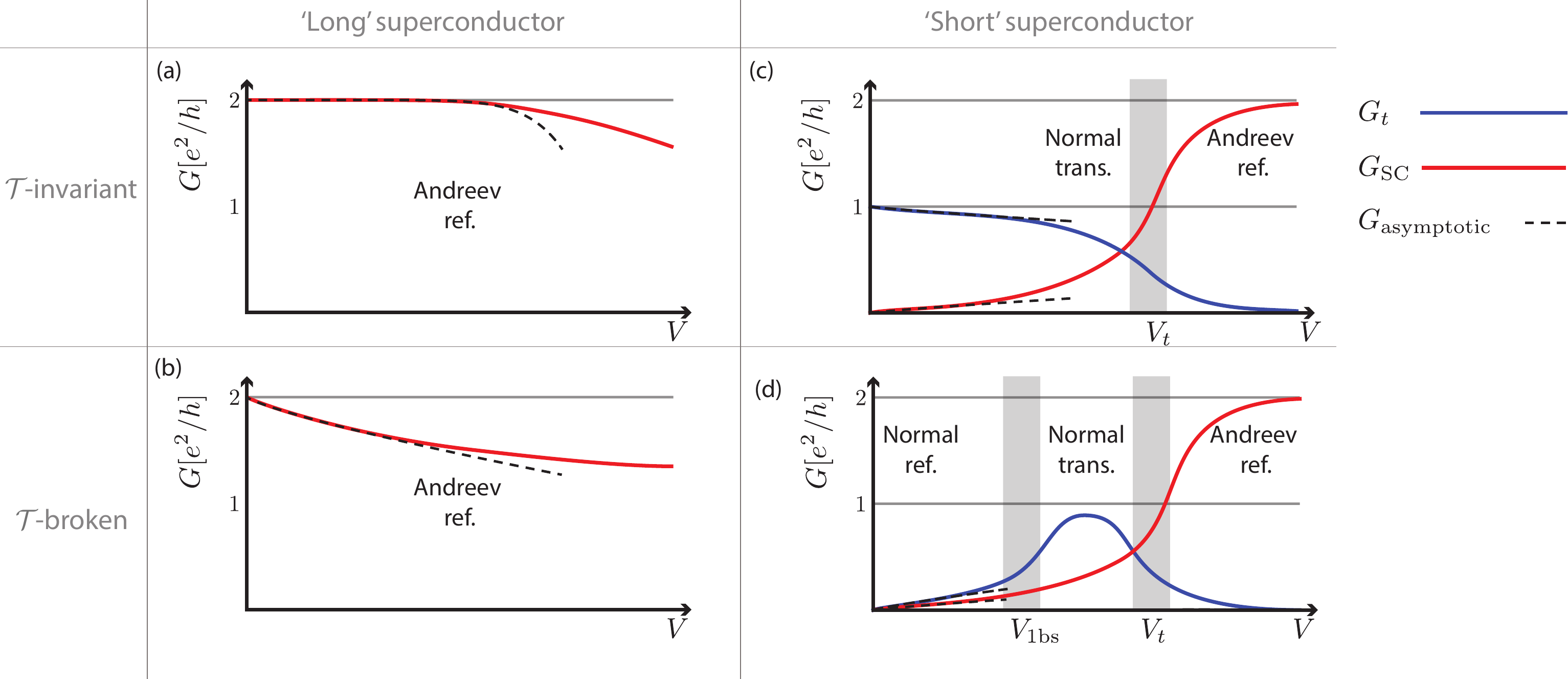}
\caption{Conductance versus bias voltage $V$ for the grounded-superconductor setup in Fig.~\ref{Device}, assuming moderate repulsive interactions in the edge-state leads.   In all panels $V$ is smaller than the induced superconducting gap.  Red and blue curves illustrate qualitative trends in the conductances $G_{\rm SC}$ and $G_t$.  Dashed lines are universal power-laws, shown for $g = 0.7$, describing the approach to quantized values as $V \rightarrow 0$ (intermediate-voltage power laws described in the text are suppressed for simplicity).  (a) and (b) Tunneling into a long superconductor yields perfect Andreev reflection and $G_{\rm SC} = 2e^2/h$ asymptotically, with stronger power-law corrections when time-reversal symmetry $\mathcal{T}$ is broken.  (c) For a short, $\mathcal{T}$-invariant superconductor perfect normal transmission sets in so that $G_{\rm SC} = 0$ and $G_t = e^2/h$ at low energies.  (d) Breaking $\mathcal{T}$ instead yields perfect normal reflection and hence vanishing conductances as $V\rightarrow 0$.  The non-monotonic $G_t$ is especially noteworthy, indicating that here the system `samples' all three fixed points as the voltage is reduced.  }   
\label{ConductancePlots}
\end{figure*}

Evidently breaking time-reversal symmetry yields only quantitative effects on the conductance in the long-superconductor limit.  Most notably, for the weak repulsive interactions assumed in this section, the voltage exponent for the $\mathcal{T}$-invariant case is quite large---reflecting strong irrelevance of two-particle backscattering at the Andreev fixed point.  Figures~\ref{ConductancePlots}(a) and (b) sketch $G_{\rm SC}$ for the $\mathcal{T}$-invariant and $\mathcal{T}$-broken settings.

%%%%%%%%%%%%%%%%%%%%%%%%%%%%%%%%%%%%%%%%%%%%%%%%

\subsection{Short-superconductor limit}

By contrast, for a short superconductor explicitly breaking $\mathcal{T}$ modifies the conductance more drastically.  Thus it will be useful to separately treat the cases with and without time-reversal symmetry.

%%%%%%%%%%%%%%%%%%%%%%%%%%%%%%%%%%%%%%%%%%%%%%%%

\subsubsection{Time-reversal-symmetric case}

Imagine beginning from a long $\mathcal{T}$-invariant superconductor with $L \gg \xi$ and then shrinking $L$ towards the short-superconductor limit.  In this thought experiment the system initially exhibits Andreev boundary conditions at each interface, but begins to develop a perturbation
\begin{equation}
  \delta H = \lambda_t \cos(\Theta_1-\Theta_2 + \chi)
  \label{deltaHt}
\end{equation}
that tunnels electrons directly past the superconductor.  Such events clearly promote non-zero $G_t$ and suppress $G_{\rm SC}$.  We will assume that the bare coupling $\lambda_t$ is weak compared to the induced Cooper-pairing gap.  At intermediate energies (e.g., when probing the system at voltages that are large compared to $\lambda_t$ but still small relative to the pairing gap) the system then to a good approximation begins at the Andreev$\oplus$Andreev fixed point, perturbed by Eq.~\eqref{deltaHt}.  Since the perturbation is relevant, $\lambda_t$ asymptotically generates perfect normal transmission boundary conditions at the lowest energy scales---yielding completely different conductances.  We are interested in predicting universal transport characteristics in both the intermediate- and low-voltage regimes as defined here.

Viewing $\lambda_t$ as a perturbation to the Andreev$\oplus$Andreev fixed point, the renormalized coupling at energy scale $E$ takes the form of Eq.~\eqref{lambdaE} with scaling dimension $\Delta_t = g$.  Since the conductance corrections are again proportional to the coupling squared, we obtain the intermediate-voltage scaling relations
\begin{eqnarray}
  \begin{cases}
  G_{\rm SC}(V) = \frac{2e^2}{h}\left[1-(V_t/V)^{2(1-g)}\right]
  \\
  G_{t}(V) = \frac{e^2}{h}(V_t/V)^{2(1-g)}
  \end{cases}
  ({\rm intermediate~}V).
  \nonumber \\
  \label{Gintermediate}
\end{eqnarray}
These relations apply when $V \gg V_t$ with some non-universal $V_t$ determined by the bare coupling strength $\lambda_t$.

To attack the low-voltage regime, we sit at the perfect normal transmission fixed point and perturb with a local Andreev-reflection term
\begin{equation}
  \delta H = \lambda_A \cos(2\Phi)
\end{equation}
that contributes a non-zero $G_{\rm SC}$ and suppresses $G_t$ below $e^2/h$.  (In Sec.~\ref{NormalTransmissionStability} we also considered perturbing the normal transmission fixed point with two-particle backscattering, but we neglect such processes here since they are much more irrelevant compared to $\lambda_A$.)  The $\cos(2\Phi)$ perturbation exhibits scaling dimension $g^{-1}$; we therefore get
\begin{eqnarray}
  \begin{cases}
  G_{\rm SC}(V) = \frac{2e^2}{h}(V/V_A)^{2(g^{-1}-1)}
  \\
  G_{t}(V) = \frac{e^2}{h}\left[1-(V/V_A)^{2(g^{-1}-1)}\right]
  \end{cases}
  ({\rm low~}V).
  \label{Glow}
\end{eqnarray}
in the low-voltage regime $V \ll V_A$, where $V_A$ follows from the bare coupling $\lambda_A$.

Equations~\eqref{Gintermediate} and \eqref{Glow} spotlight the nontrivial impact that even weak repulsive interactions have on transport in the short-superconductor case.  Indeed, if we fine-tune to the non-interacting limit $g = 1$, the voltage dependence drops out of these expressions; the power-law forms should then be replaced by non-universal, constant conductance corrections as found in the free-fermion treatment from Ref.~\onlinecite{NonInteractingEdgeStateConductance}.  The non-universality at $g = 1$ is symptomatic of the fact that the free-fermion problem sits at the phase boundary between two different stable boundary conditions; recall Fig.~\ref{PhaseDiagram}(c).  As a corollary, with weak repulsive interactions the nontrivial power-law corrections in Eqs.~\eqref{Gintermediate} and \eqref{Glow} turn on rapidly as the voltage changes, contrary to the long superconductor where the voltage corrections in Eq.~\eqref{GSClong} come with much larger exponents for $g$ slightly below 1.  Figure~\ref{ConductancePlots}(c) sketches the predicted conductances $G_{\rm SC}$ and $G_t$ for the short, $\mathcal{T}$-invariant superconductor setup.

%%%%%%%%%%%%%%%%%%%%%%%%%%%%%%%%%%%%%%%%%%%%%%%%

\subsubsection{Time-reversal-broken case}

Breaking time-reversal generates single-electron backscattering events that destabilize the normal transmission fixed point in favor of perfect normal reflection at each end of the superconductor.  We will assume that the backscattering amplitude $\lambda_\text{1bs}$ is weak compared to the bare tunneling strength $\lambda_t$, which can always be arranged in practice (e.g., by using weak magnetic fields).  In this case the system remains well described by the perfect normal transmission fixed point over an extended energy (and voltage $V_\text{1bs}\ll V\ll V_t$) window, with normal reflection kicking in only at the lowest energy scales $V\lesssim V_\text{1bs}$.  Following the preceding analysis, we will explore how the conductances evolve as the system initially flows away from the normal transmission fixed point, and then upon approaching the stable normal$\oplus$normal fixed point.

At the normal transmission fixed point, the single-electron backscattering perturbation reads
\begin{equation}
  \delta H = \lambda_{\rm 1bs} \cos(2\Theta)
\end{equation}
and carries a scaling dimension of $g$.  The correspondingly reduced conductance $G_t$ thus scales according to
\begin{equation}
  G_t = \frac{e^2}{h}\left[1-(V_{\rm 1bs}/V)^{2(1-g)}\right], ~~~({\rm intermediate}~V)
  \label{Gintermediate_broken}
\end{equation}
over the intermediate voltage range $V_t \gg V \gg V_{\rm 1bs}$. 

Finally, for the smallest voltages we sit at the stable normal$\oplus$normal fixed point and add the leading irrelevant perturbations---which encode electron tunneling across the superconductor and crossed-Andreev reflection,
\begin{equation}
  \delta H = \lambda_t \cos(\Phi_1-\Phi_2) + \lambda_{\rm CAR}\cos(\Phi_1 + \Phi_2).
  \label{Hperts}
\end{equation}
These two operators possess the same scaling dimension $g^{-1}$.  While only $\lambda_{\rm CAR}$ contributes to $G_{\rm SC}$, both terms nontrivially influence the transmitted current $I_t$ and hence $G_t$.  Thus we obtain
\begin{eqnarray}
  \begin{cases}
  G_{\rm SC}(V) = \frac{2e^2}{h}(V/V_{\rm CAR})^{2(g^{-1}-1)}
  \\
  G_{t}(V) = \frac{e^2}{h}(V/V_*)^{2(g^{-1}-1)}
  \end{cases}
  ({\rm low~}V),
  \label{Glow_broken}
\end{eqnarray}
for $V \ll V_{\rm CAR}, V_*$, where $V_{\rm CAR}$ is set by the bare value of $\lambda_{\rm CAR}$ while $V_*$ is a function of both couplings in Eq.~\eqref{Hperts}.

Putting these results together, we find that the conductances for a short superconductor with weakly broken $\mathcal{T}$ evolve nontrivially with voltage as shown in Fig.~\ref{ConductancePlots}(d).  The non-monotonic voltage dependence of $G_{t}$ is particularly striking; this interesting feature highlights the starkly different sensitivity to time-reversal-symmetry breaking for setups featuring a long superconductor that realizes a bona fide topological phase and a short quantum-dot-like superconductor that is in no sense topological.

%%%%%%%%%%%%%%%%%%%%%%%%%%%%%%%%%%%%%%%%%%%%%%%%

\section{Floating superconductor with charging energy}
\label{sec:FloatingSC}

\begin{figure}
	\centering
	\includegraphics[width=8.5cm]{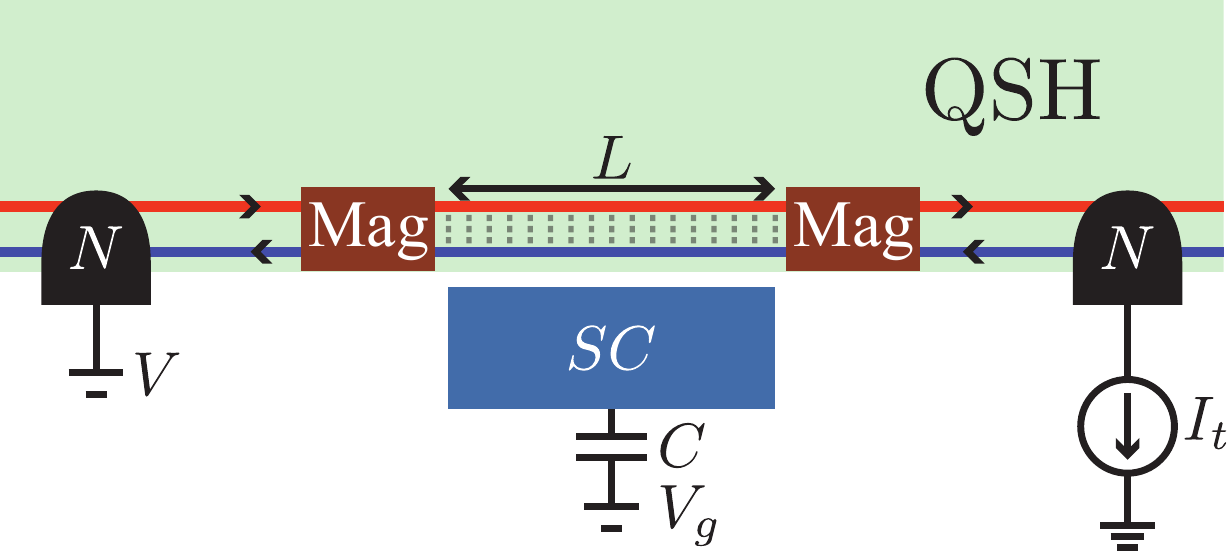}
	\caption{Quantum-spin-Hall system proximitized by a floating superconductor.  Magnetized regions on each end define a superconducting island with charging energy; the gate voltage $V_g$ tunes the island charge.  The magnetized regions are created via adjacent ferromagnetic insulators (which explicitly violate time-reversal symmetry) or two-particle backscattering at the edge (which breaks time-reversal spontaneously).  
	}
	\label{DeviceWithChargingEnergy}
\end{figure}

Next we explore transport in a QSH device proximitized by a floating---rather than grounded---superconductor.  Figure~\ref{DeviceWithChargingEnergy} depicts the specific setup of interest (for studies of related systems see Refs.~\onlinecite{LiangFuTeleport,Egger,Exponential,vanHeck}).  Weak links on each side of the Cooper-paired QSH edge modes define a superconducting island with finite charging energy $E_C$.  We consider the cases where the weak links are generated by $(i)$ ferromagnetic regions [$\cos(2\theta)$ terms] that break $\mathcal{T}$ explicitly and $(ii)$ relevant two-particle backscattering [$\cos(4\theta)$ terms] that breaks $\mathcal{T}$ spontaneously.  To single out charging over finite-size effects, throughout this section we assume that the superconducting region is much longer than the induced coherence length ($L\gg \xi$).  In the absence of charging energy the system then supports zero modes that encode robust ground-state degeneracies.  Turning on $E_C$ generically lifts these degeneracies, though imprints of the zero modes appear through nontrivial Coulomb-blockade behavior that can be probed by sending current across the island via the adjacent gapless edge states in Fig.~\ref{DeviceWithChargingEnergy}.  We will quantify this behavior by first exploring the superconducting island's ground-state charge configurations, which already reveals interesting physics, and then studying fixed points that describe the device's universal transport characteristics.

%%%%%%%%%%%%%%%%%%%%%%%%%%%%%%%%%%%%%%%%%%%%%%%%

\subsection{Charging patterns}

We phenomenologically incorporate Coulomb interactions on the superconducting island by adding a charge-dependent energy shift of $E_C (\hat{n}-n_0)^2$.  Here $\hat{n}$ counts the total number of electrons on the island (arising both from the paired edge region and parent superconductor) while $n_0(V_g)$ is a gate-tunable offset.  We are interested in quantifying how the total charge in the island's ground state varies upon sweeping the gate voltage $V_g$.  As a baseline, recall that a \emph{conventional} superconducting dot with a `large' pairing gap $\Delta$ exhibits $2e$ charge-addition periodicity, i.e., varying $V_g$ successively adds pairs of electrons to the ground state since single electrons pay an additional energy $\Delta$.  Our QSH setup, by contrast, yields a richer charge-addition pattern.  

\begin{figure}
	\includegraphics[width=8.5cm]{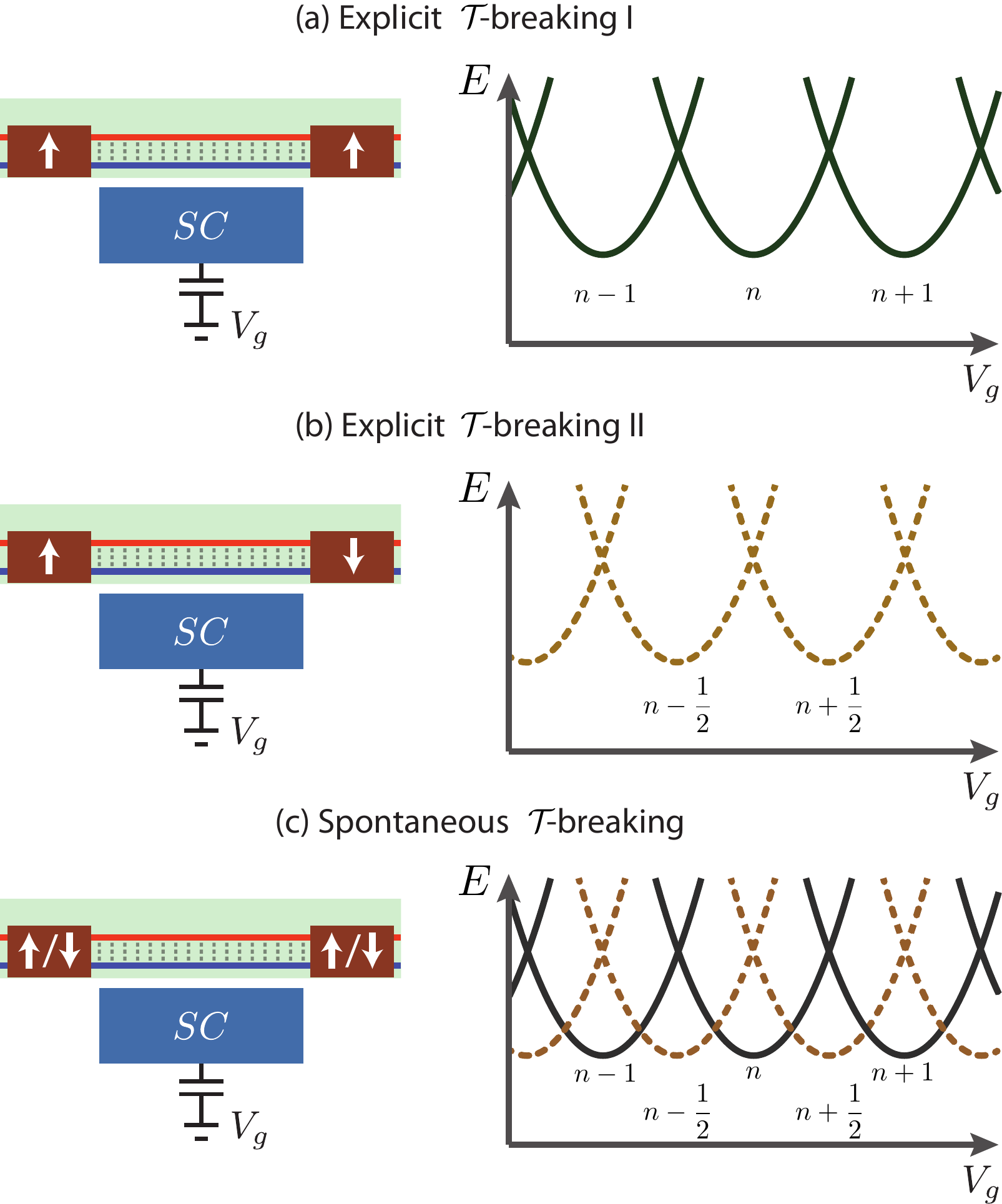}
	\caption{Left: Superconducting island bordered by (a/b) ferromagnetic regions with parallel/antiparallel magnetizations and (c) regions with spontaneously broken time-reversal symmetry driven by interactions.  Right: Corresponding low-lying energy levels versus gate voltage $V_g$ for different island charge states.  Parabolas are labeled by the total electron number, where $n$ is an integer.  When time-reversal symmetry is explicitly broken [cases (a) and (b)] sweeping $V_g$ adds electrons one at a time to the island, with an interesting half-integer offset in the case of antiparallel magnetizations.  With spontaneously broken time-reversal symmetry [case (c)] charges are instead added in $e/2$ increments.   These unusual $e$ and $e/2$ charge-addition periodicities are remnants of Majorana and parafermion zero modes present when the superconductor is grounded.  	
	}
	\label{ChargingEnergyFig}
\end{figure}

Let us begin with the explicit $\mathcal{T}$-broken case where ferromagnets bordering the superconductor polarize along the same direction; see Fig.~\ref{ChargingEnergyFig}(a), left side.  The right side of Fig.~\ref{ChargingEnergyFig}(a) sketches the low-lying energies versus gate voltage, with each parabola representing a different integer electron number ($\cdots, n-1, n,n+1, \cdots$) for the island.  Note the lack of offset for even- versus odd-charge parabolas: the latter states no longer incur a pairing-energy penalty, manifesting the Majorana zero modes present when the superconductor is grounded.  Sweeping $V_g$ thus clearly adds electrons to the ground state one at a time (rather than in pairs) as first predicted in important work by Fu\cite{LiangFuTeleport}.

It is useful to recover this conclusion in bosonized language.  Both ferromagnets generate an identical perturbation $\lambda_{\rm FM} \cos(2\theta)$ that locally gaps the edge by pinning $\theta$ to a minimum of the cosine.  The difference between the pinned values across the superconductor is $\Delta \theta = q \pi$ for some integer $q$.  Physically, $q$ is the charge on the intervening edge segment---which is conserved mod 2---since $\Delta \theta = \int_x \partial_x\theta = \pi \int_x \rho$ ($\rho$ is the edge density).  States with $q$ even and odd respectively carry even and odd total island charge, and are distinguished energetically by charging energy but no other terms in the Hamiltonian.  This analysis is consistent with the energies sketched in Fig.~\ref{ChargingEnergyFig}(a).  

Suppose that we now rotate the ferromagnet on the right such that its magnetization orients antiparallel to that on the left, producing the configuration in Fig.~\ref{ChargingEnergyFig}(b).  The gap-opening perturbation under the right ferromagnet then acquires an overall minus sign and reads $-\lambda_{\rm FM}\cos(2\theta)$ (one way to see this is to recall that time reversal sends $\theta \rightarrow \theta+\pi/2$).  We thus obtain $\Delta \theta = q \pi$ with \emph{half-integer} $q$.  In other words, the magnetization reversal pumps an $e/2$ fractional charge onto the island,\cite{qi_fractional_2008} yielding the shifted energy curves in Fig.~\ref{ChargingEnergyFig}(b) that are labeled by half-integer electron numbers ($\cdots, n-1/2, n+1/2, \cdots$) for the island.  Sweeping the gate voltage again adds electrons one at time to the ground state, which however now possesses a nontrivial fractional offset charge.\footnote{This offset charge need not be half-integer, and instead varies continuously upon changing the relative orientation of the ferromagnets' magnetizations, as in Ref.~\onlinecite{qi_fractional_2008}.}  

Reference~\onlinecite{qi_fractional_2008} in fact predicted identical charging patterns, including the fractional offset for antiparallel magnetizations, for a \emph{non-superconducting} island created by ferromagnetic domains.  Cooper pairing elevates the gap within each charge sector from $\sim 1/L$ to of order the pairing energy but does not alter the charging periodicity---a nontrivial property that sharply distinguishes the system from a conventional superconducting dot.  

Replacing the ferromagnets by spontaneously magnetized regions [see Fig.~\ref{ChargingEnergyFig}(c)] allows the system to access both the integer \emph{and} half-integer charge sectors simultaneously.  In this setup the superconducting island is created by gap-opening perturbations $\lambda_{\rm 2bs}\cos(4\theta)$ that result in $\Delta \theta = \pi q$, where $q$ can be either integer or half-integer.  The energies versus gate voltage thus appear as shown in Fig.~\ref{ChargingEnergyFig}(c), implying $e/2$ charge-addition periodicity for an island with spontaneously broken time-reversal symmetry.  This result reflects the parafermion zero modes that appear with a grounded superconductor, and can be readily understood from Figs.~\ref{ChargingEnergyFig}(a) and (b): On sweeping the gate voltage, the weak links can now flip their magnetizations dynamically to minimize charging energy since the `up' and `down' orientations are on equal footing.  By contrast, the edge magnetizations in Figs.~\ref{ChargingEnergyFig}(a) and (b) are slaved to the adjacent ferromagnets and thus cannot flip without paying a large energy, thereby halving the number of accessible charge states.  

The charging patterns identified above are essential for understanding conduction across the superconducting island, which we study in the remainder of this section.  For a given setup (i.e., explicit versus spontaneous time-reversal breaking) there are two cases to consider: `off-resonant' transport corresponding to generic gate voltages that yield a unique lowest-energy charge configuration for the island, and `on-resonant' transport where degeneracies arise because $V_g$ is fine-tuned to a crossing between adjacent parabolas in Fig.~\ref{ChargingEnergyFig}.  The analysis is simplified by the absence of Andreev processes, which are frozen out because the parent superconductor can no longer absorb Cooper pairs with impunity.  Thus the only physical fixed points describe perfect normal reflection at each end of the island and perfect normal transmission (perfect Andreev reflection would lead to a constant rate of accumulating or depleting Cooper pairs in the floating superconductor).  To stay within the picture of a well-defined island we will start at the normal$\oplus$normal fixed point and study its stability towards transmitting perturbations. Since there are only two available fixed points it is reasonable to assume that the system flows towards perfect normal transmission once normal reflection becomes unstable.

%%%%%%%%%%%%%%%%%%%%%%%%%%%%%%%%%%%%%%%%%%%%%%%%

\subsection{Island with explicitly broken time-reversal symmetry}

In the absence of interactions outside of the island, tunneling across a superconducting island defined by ferromagnetic barriers was first studied by Fu.\cite{LiangFuTeleport}  The physics is essentially described by a single-level quantum dot ($f$) with energy $\varepsilon$ coupled to the left and right QSH edge modes with strength $t_1$ and $t_2$,
\begin{equation}
	H= H_{\rm lead}+\varepsilon f^{\dagger}f + \left[(t_1 \psi_1^\dagger-it_2 \psi_2^\dagger)f +{\rm H.c.}\right].
\end{equation}
Here $\psi_{1,2}$ denote gapless edge fields evaluated at the left/right side of the island while the dot level crudely models the lowest two charge states of the island [e.g., $f^\dagger f = 0$ and $1$ represent charge states $n$ and $n+1$ in Fig.~\ref{ChargingEnergyFig}(a) with $\varepsilon$ their energy difference at a particular gate voltage].  Note that this Hamiltonian holds independent of any offset charge in the island's ground state; the following discussion thus applies to both the parallel and antiparallel magnetizations displayed in Figs.~\ref{ChargingEnergyFig}(a) and (b).  

\begin{figure}
	\centering
	\includegraphics[width=8.5cm]{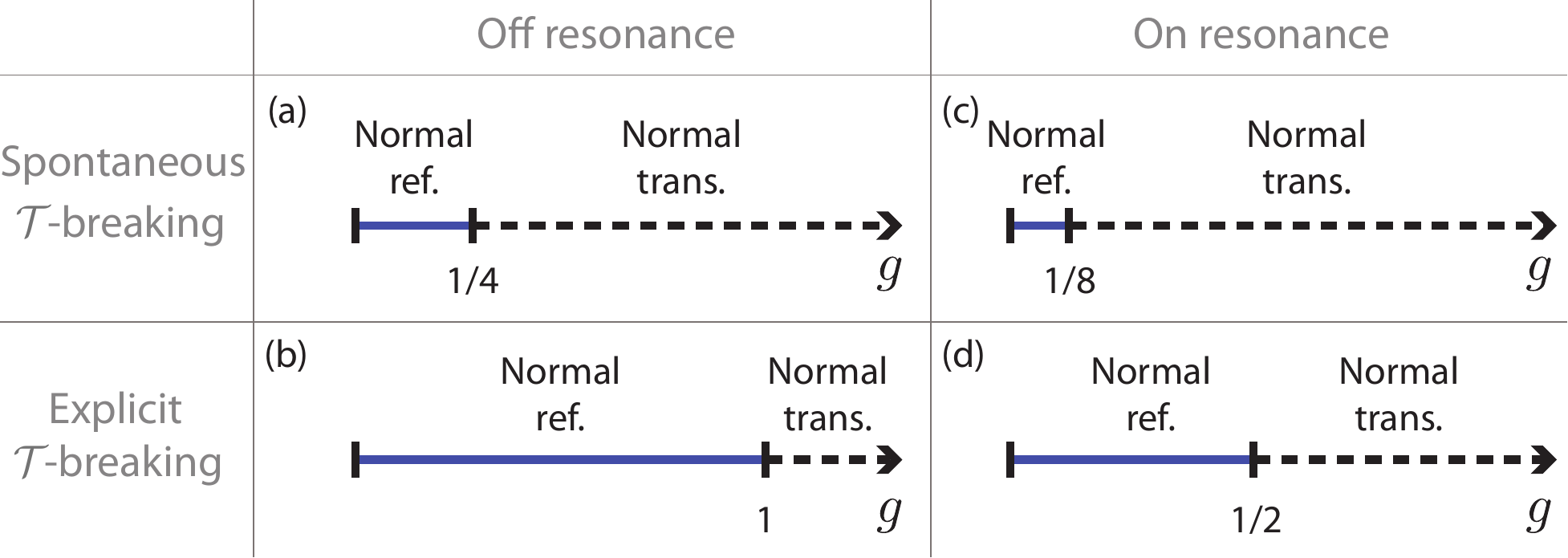}
	\caption{Phase diagrams for the floating-superconducting setup (Fig.~\ref{DeviceWithChargingEnergy}) as a function of the Luttinger parameter $g$ in the adjacent gapless helical edges.  (The on-resonance column corresponds to symmetric barriers.)  Since charging energy freezes out Andreev processes, the only available fixed points are perfect normal transmission and perfect normal reflection that respectively yield conductances $G_t = e^2/h$ and $0$.  The off-resonance cases [(a) and (b)] arise for generic gate voltages $V_g$ at which adding charge to the floating superconductor costs finite charging energy.  Fine-tuning $V_g$ to degeneracies between different charge states---i.e., crossings between parabolas in Fig.~\ref{ChargingEnergyFig}---produces the on-resonance cases [(c) and (d)]. }
	\label{ChargingEnergyPhaseDiagram}
\end{figure}

In the off-resonant case ($\varepsilon\neq0$), the single level can be integrated out yielding an effective coupling that tunnels an electron between the left and right leads, $\sim \frac{t_1t_2}{\varepsilon}\cos(\Phi_1-\Phi_2)$ in bosonized language.  The discussion then becomes similar to that of Sec.~\ref{sec:normal_normal_broken}. Moreover, due to the absence of Andreev processes the solution of this problem is well know in the literature of 1D systems \cite{Kane-Fisher-PRB}. For repulsive interactions $g<1$ the normal$\oplus$normal fixed point is stable while for $g>1$ the system flows to perfect transmission as summarized in Fig.~\ref{ChargingEnergyPhaseDiagram}(b).

On resonance ($\varepsilon = 0$), the system is best described by a Coulomb gas model \cite{KaneFisherResonantTunneling,Kane-Fisher-PRB} with charges $e$ hopping from the left and right leads on and off the island. For symmetric couplings $t_1=t_2$ these resonant hoppings lead to perfect normal transmission for attractive or not-too-strong repulsive interactions $g>1/2$ and perfect normal reflection at $g<1/2$ [see Fig.~\ref{ChargingEnergyPhaseDiagram}(d)].  Perfect transmission in the former regime has been referred to as electron teleportation \cite{LiangFuTeleport} because the effect is facilitated by Majorana modes that provide a single non-local fermionic level even for long islands with $L\gg \xi$. Note, however, that with repulsively interacting leads ($g<1$) perfect transmission only happens for symmetric couplings $t_1=t_2$. For asymmetric couplings one recovers the off-resonant phase diagram \cite{Kane-Fisher-PRB} and therefore perfect normal reflection for any $g<1$.

In summary, with sufficiently weak repulsive interactions in the leads, sweeping the gate voltage on an island defined by symmetric ferromagnetic barriers yields $e$-periodic Coulomb-blockade peaks with zero-bias conductance $G_t$ that asymptotically approaches $e^2/h$ on resonance and vanishes in between peaks.

%%%%%%%%%%%%%%%%%%%%%%%%%%%%%%%%%%%%%%%%%%%%%%%%

\subsection{Island with spontaneously broken time-reversal symmetry}

The preceding arguments underlying Coulomb blockade physics apply similarly to a superconducting island with weak links that spontaneously break $\mathcal{T}$ symmetry.  Interestingly, however, transport in this case is mediated by transfer of fractional charges across the island.  By inspecting Fig.~\ref{ChargingEnergyFig}(c) we indeed see that the energetically cheapest way to pass current between the leads (at any gate voltage) involves incrementing the island charge by $e/2$.  Off resonance, such processes can be treated perturbatively and yield an effective coupling $\sim \cos\left(\frac{\Phi_1-\Phi_2}{2}\right)$ of the same form that promotes normal transmission for the grounded short-superconductor setup [see Eq.~\eqref{NNperturbation}].  Thus in the off-resonant regime the main effect of charging energy is to couple the leads via the non-local parafermion modes on the island. The critical interaction strength at which the normal-reflection fixed point becomes unstable towards perfect transmission can be read off from Sec.~\ref{sec:normal_normal_T} and is given by $g=1/4$; Fig.~\ref{ChargingEnergyPhaseDiagram}(b) illustrates the phase diagram.  

The $g = 1/4$ phase boundary can alternatively be obtained from a straightforward generalization of Kane and Fisher's analysis \cite{Kane-Fisher-PRB} of an off-resonant quantum dot in a Luttinger liquid. The latter can also be described by a Coulomb gas model (slightly different form that of the resonant case) where charges hop from the left to the right lead. In the presence of parafermions these charges have a value $ke$ with $k=1/2$. The Coulomb gas model is expressed in terms of logarithmically interacting charges with an interaction strength proportional to $1/g$. We can therefore deduce the phase diagram for general $k$ by using the $k=1$ result and renormalizing $g\rightarrow g/k^2$ (the interaction is quadratic in the charges). Since for $k=1$ the phase boundary occurs at $g=1$, halving the charge yields $g = 1/4$ as found above.  

By the same argument we can also immediately deduce the on-resonance phase diagram with symmetric barriers from the results of Kane and Fisher \cite{Kane-Fisher-PRB}. The phase boundary shifts to $g=1/2$ in the case of resonant electron tunneling, implying a phase boundary of $g=1/8$ in our setup with resonant $e/2$ tunneling; see Fig.~\ref{ChargingEnergyPhaseDiagram}(c).  With asymmetric barriers, the system again follows the off-resonant phase diagram\cite{Kane-Fisher-PRB}.  

The results for an island created by spontaneous $\mathcal{T}$ breaking differ quite dramatically from the ferromagnetic-barrier setup discussed in the previous subsection.  Notably, the enhanced stability window for perfect normal transmission implies that the conductance $G_t$ asymptotically approaches $e^2/h$ for \emph{arbitrary} gate voltages whenever $g>1/4$---which includes the most experimentally relevant case of weak interactions in the edge-state leads.  Anomalous $e/2$-periodic  Coulomb-blockade peaks are visible (again asymptotically) only in the restricted window $1/8 < g < 1/4$ within which normal reflection is stable off resonance but unstable on resonance.  Nevertheless, we expect that even with a weakly interacting lead the finite-temperature/voltage conductance reveals signatures of $e/2$ charging physics 
since the corrections to the quantized conductance will differ on and off resonance.  It would be interesting in future work to quantify the gate-voltage dependence of the conductance in this case.  

\subsection{Comparison to the grounded-superconductor case}

Comparing Fig.~\ref{PhaseDiagram} with Fig.~\ref{ChargingEnergyPhaseDiagram} reveals striking similarities between the phase diagrams of the floating- and grounded-superconductor setups. Specifically, the phase boundary of the normal-reflecting region of the long (short) grounded superconductor coincides with that of the resonant (off-resonant) floating superconductor. This agreement is actually rather natural: A hybridization-split zero mode in a short grounded superconductor acts as a finite-energy off-resonant level (and vice versa, as the off-resonant level introduces an effective tunneling between the ends of the superconductor). The long grounded superconductor case can be seen as resonant (Andreev) tunneling from the wire back to itself via a single zero mode. Particle hole-symmetry then ensures that the tunnel couplings of particles and holes in the wire to the zero mode are exactly the same, thus guaranteeing resonant tunneling \cite{law_2009}.

In light of this similarity one might wonder how the local resonant (Andreev) tunneling and the non-local resonant (normal) tunneling regimes connect in the limit $E_C\rightarrow 0$. Interestingly, the two-terminal conductance is $e^2/h$ in both regimes\cite{ulrich_2015}. In the Andreev tunneling regime this result follows from adding the independent resistances $(2e^2/h)^{-1}$ at both sides of the (long) wire. The crucial effect of charging energy is therefore to provide coherence for the resonant tunneling through the superconductor \cite{ulrich_2015}. One should however note that this coherence will be lost for temperatures $T>E_C$ (where multiple levels contribute). Since $E_C\propto 1/L$ the `teleportation' is therefore 
reminiscent of resonant tunneling through a quantum dot made out of a normal 1D wire, where the level spacing ($\propto 1/L$) gives a similar condition for the temperature.

%%%%%%%%%%%%%%%%%%%%%%%%%%%%%%%%%%%%%%%%%%%%%

\section{Conclusions}
\label{sec:conclusions}

In this paper we explored the transport characteristics of QSH architectures that support time-reversal-invariant topological superconductivity with no analog in purely 1D systems.  Our analysis accordingly uncovered numerous sharp distinctions from analogous nanowire-based devices with explicitly broken $\mathcal{T}$:

$(i)$ For the long $(L \gg \xi)$ grounded-superconductor setup, $\mathcal{T}$ symmetry restricts the allowed backscattering processes and thus further promotes Andreev reflection relative to strict 1D geometries.  Universal power-law corrections to the quantized zero-bias conductance are suppressed significantly; moreover, the perfect-Andreev-reflection fixed point remains stable down to much stronger repulsive interactions in the leads---the critical value becomes $g = 1/8$ instead of $1/2$.  

$(ii)$ At $g<1/8$ time-reversal symmetry is broken \emph{spontaneously}, yielding perfect normal reflection at low energies.  Qualitative differences from nanowires nevertheless persist.  In particular, this strongly interacting regime is most profitably viewed in terms of hybridization between the edge states and a dynamically generated parafermion zero mode.  As a technical aside, we note that without this viewpoint the consistency between the flows at the normal and Andreev fixed points becomes greatly obscured.  

$(iii)$ Without interactions the short $(L \lesssim \xi)$ ground-superconductor setup exhibits non-universal transport; in renormalization-group language the system sits at the boundary between two stable fixed points.  Arbitrarily weak repulsive interactions drive a flow to a fixed point characterized by perfect transmission across the short superconductor.  The stability of perfect normal transmission arises from the backscattering restriction imposed by $\mathcal{T}$ symmetry, and is destroyed in favor of normal reflection only with strong interactions $(g<1/4)$.  By contrast, for a nanowire stable perfect transmission is absent entirely, as the superconductor instead `cuts' the wire for any $g<1$.  

$(iv)$ Upon explicitly breaking $\mathcal{T}$ perfect transmission similarly disappears for the QSH edge.  The exquisite sensitivity to $\mathcal{T}$-breaking perturbations
underlies nontrivial transport predictions summarized in Figs.~\ref{ConductancePlots}(c,d).  Weak $\mathcal{T}$ breaking thereby provides a handy experimental knob for contrasting to the bona fide topological long-superconductor limit where Andreev reflection tends to dominate with or without $\mathcal{T}$; cf.~Figs.~\ref{ConductancePlots}(a,b).  

$(v)$ Creating a floating superconducting island with charging energy at the QSH edge requires breaking $\mathcal{T}$ either explicitly or spontaneously via strong interactions to isolate the paired region.  Both methods offer interesting extensions to Coulomb-blockade physics in nanowire counterparts.  With explicit $\mathcal{T}$ breaking it becomes possible to trap a fractional offset charge on the island by twisting the relative orientation of the barrier magnetizations.  Spontaneous $\mathcal{T}$ breaking allows the fractional offset charge to switch \emph{dynamically}, leading to a novel $e/2$-periodic charging pattern for the island that originates from parafermion modes.

It is worth emphasizing that all of the above results require proximitized \emph{helical} QSH edge modes.  Notably, `accidental' edge states such as those identified in Ref.~\onlinecite{Nichele} would yield only trivial superconductivity under similar conditions and thus exhibit entirely different behavior.  Testing our predictions for the grounded-superconductor setups appears particularly accessible for experiments given the minimal ingredients required---a QSH system with an inert bulk, superconducting proximity effect, and weak magnetic fields.   Verifying even the qualitative trends that we identified would provide valuable insight into the unique brand of topological superconductivity possible in this setting, and perhaps also provide further evidence for the helical nature of the edge modes themselves.  Pursuing islands with charging energy appears more challenging due to the requirement of introducing magnetic barriers.  Devising practical alternative realizations for such fractional Coulomb-blockade physics poses an interesting challenge for future research.

%%%%%%%%%%%%%%%%%%%%%%%%%%%%%%%%%%%%%%%%%%%%%%%%%%%%%%%%%%%%%%%%%%%%%%%%%%%%%%

\acknowledgments{
We thank Debaleena Nandi, Roman Lutchyn and Amir Yacoby for illuminating discussions.  We also gratefully acknowledge support from the National Science Foundation through grant DMR-1341822 (D.~A., S.-P.~L., and J.~A.); the NSERC PGSD program (D.~A.); the Caltech Institute for Quantum Information and Matter, an NSF Physics Frontiers Center with support of the Gordon and Betty Moore Foundation through Grant GBMF1250; and the Walter Burke Institute for Theoretical Physics at Caltech.
}

\appendix

%%%%%%%%%%%%%%%%%%%%%%%%%%%%%%%%%%%%%%%%%%%%%%%%

\section{Duality for perfect-Andreev-reflection fixed point perturbed by two-particle backscattering
\label{APP:duality}}

In this Appendix we apply a duality transformation to understand the boundary fixed points and perturbations for a strongly interacting gapless edge that impinges on a `long' superconductor.  We begin with the partition function at the perfect-Andreev-reflection fixed point perturbed by $\cos{(4\Theta)}$,
\begin{equation}
Z = \int \mathcal{D} \Theta e^{-S_A[\Theta] + 2\lambda_{{\rm 2bs}} \int d\tau \cos{(4 \Theta)}}
\end{equation}
with $\lambda_{{\rm 2bs}}$ positive for concreteness (the prefactor of 2 is inserted for convenience).  As discussed in Sec.~\ref{AndreevStability1}, $\lambda_{{\rm 2bs}}$ is relevant for $g<1/8$---which we assume here---and destabilizes the Andreev boundary conditions.  Duality provides a useful viewpoint on the system's fate under renormalization.

We first employ the Villain approximation for the cosine,
\begin{equation}
e^{2\lambda_{{\rm 2bs}} \cos{4\Theta}} \rightarrow e^{2\lambda_{{\rm 2bs}}}\sum_{n(\tau) \in \mathbb{Z}} e^{-\lambda_{{\rm 2bs}} (4\Theta-2\pi n)^2},
\end{equation}
and then introduce a Hubbard-Stratonovitch field $\rho(\tau)$ to decouple the quadratic term,
\begin{eqnarray}
&& e^{-\lambda_{{\rm 2bs}} (4\Theta - 2\pi n)^2} =
\nonumber \\
&& \int \mathcal{D}\rho e^{-\int d \tau [\rho^2/\lambda_{{\rm 2bs}} + 2i \rho (4\Theta - 2\pi n)]}.
\end{eqnarray}
Putting these together and discarding unimportant constants, we obtain
\begin{equation}
Z = \int \mathcal{D} \Theta \mathcal{D} \rho \sum_{n(\tau) \in \mathbb{Z}} e^{-S_A[\Theta] - \int d \tau [\rho^2/\lambda_{{\rm 2bs}} + 2 i \rho (4\Theta - 2\pi n)]}.
\end{equation}

Next we write $\rho = \partial_{\tau} \Phi/(8\pi)$; the $\rho\Theta$ term in the above action then implies that $\Phi/\pi$ is conjugate to $\Theta$.  In these variables the partition function becomes
\begin{eqnarray}
Z &=& \int \mathcal{D}\Theta \mathcal{D}\Phi \sum_{n(\tau)\in\mathbb{Z}}
\nonumber \\
&\times& e ^{-S_A[\Theta] - \int d\tau \left[\frac{1}{\lambda_{{\rm 2bs}}}\left(\frac{\partial_\tau \Phi}{8\pi}\right)^2 + i\frac{\partial_\tau \Phi}{4\pi}(4\Theta - 2\pi n)\right]}
\end{eqnarray}
Summing over $n(\tau)$ restricts $\Phi(\tau)$ to integer multiples of $4\pi$.  We enforce this constraint `softly' by adding a $-v\cos{(\Phi/2)}$ term to the action, with $v>0$ so that $\Phi(\tau) \in 4\pi \mathbb{Z}$ is favored energetically.  Integrating out $\Theta$ then yields a partition function expressed solely in terms of $\Phi$:
\begin{equation}
  Z = \int \mathcal{D} \Phi e^{-\int \frac{d \omega}{2\pi} \frac{g|\omega|}{2\pi} |\Phi_\omega|^2 - \int d\tau \left[ \frac{1}{\lambda_{{\rm 2bs}}}\left(\frac{\partial_\tau \Phi}{8\pi} \right)^2 - v \cos{(\Phi/2)} \right]}.
\end{equation}
The $(\partial_\tau \Phi)^2$ piece is irrelevant compared to the $|\omega||\Phi_\omega|^2$ term and thus may be safely discarded when exploring low-energy behavior.  We thus obtain the desired form for the partition function,
\begin{equation}
  Z = \int \mathcal{D} \Phi e^{-S_{\rm dual}},
\end{equation}
expressed in terms of the dual action
\begin{equation}
  S_{\rm dual} = \int \frac{d \omega}{2\pi} \frac{g|\omega|}{2\pi} |\Phi_\omega|^2 - v \int d \tau \cos{(\Phi/2)}.
\end{equation}

The first term exactly reproduces the perfect-normal-reflection fixed-point action [recall Eq.~\eqref{NormalReflectionFixedPointAction}], while the second is the dual counterpart of the $\cos(4\Theta)$ perturbation that destabilizes the Andreev boundary conditions when $g<1/8$.  Since $\cos(\Phi/2)$ is irrelevant over that same range of $g$, the duality analysis strongly hints that the system flows to the stable normal-reflection fixed point with the $v$ term comprising the leading perturbation.  The following Appendix further substantiates this conclusion by deriving the $\cos(\Phi/2)$ perturbation from a more microscopic treatment.

%%%%%%%%%%%%%%%%%%%%%%%%%%%%%%%%%%%%%%%%%%%%%%%%

\section{Parafermion zero mode hybridization}
\label{APP:Parafermion}

We now revisit the geometry in Fig.~\ref{Parafermion} that supports adjacent domains gapped by two-particle backscattering [i.e., $\cos(4\theta)$] and superconductivity [i.e., $\sin(2\varphi)$].  These regions respectively favor pinning $\theta = \pi \hat n_\theta/2$ and $\varphi = \pi(\hat n_{\varphi}  + 1/4)$, where $\hat n_{\varphi},\hat n_{\theta}$ are integer-valued operators that distinguish different minima of the cosine and sine potentials.  Equation~\eqref{eq:time_reversal} implies that time-reversal transforms these operators as
\begin{equation}
  \mathcal{T}[\hat n_\theta] = \hat n_{\theta} + 1,~~~~~ \mathcal{T}[\hat n_\varphi] = -\hat n_{\varphi}-1.
  \label{Tns}
\end{equation}
Crucially, the commutations relations between $\varphi$ and $\theta$ in turn yield the nontrivial commutator $[\hat n_\varphi, n_\theta] = 2i/\pi$; thus the integer operators can \emph{not} take on well-defined eigenvalues simultaneously.  In a basis where $\hat{n}_\varphi$ is diagonal, $\hat n_\theta$ fluctuates and vice versa.

One can define a $\mathbb{Z}_4$ parafermion-zero-mode operator,\cite{ZhangKane,parafermions}
\begin{equation}
  \alpha = e^{i (\pi/2)(\hat n_{\varphi} + \hat n_{\theta})},
  \label{alphadef}
\end{equation}
that cycles between adjacent minima of the potentials.  Microscopically, Eq.~\eqref{alphadef} emerges upon projecting $e^{i (\varphi/2 + \theta)}$, evaluated in the domain wall, into the low-energy sector for the adjacent gapped regions.  It is worth emphasizing that $\alpha$ by itself is not a physical, gauge-invariant operator.  Physical perturbations involving $\alpha$ do, however, arise from hybridization between the adjacent gapless edge and the domain wall.  In particular, consider the operator
\begin{equation}
  \mathcal{O} \equiv \alpha e^{-i (\Phi/2 + \Theta)} + H.c.,
  \label{O}
\end{equation}
where $\Phi, \Theta$ continue to label bosonized fields $\varphi,\theta$ acting at the boundary of the gapless region.  Equation~\eqref{O} can be expressed solely in terms of currents and densities [i.e., $\mathcal{O} \propto e^{i\int_x (\partial_x\varphi/2 + \partial_x \theta)}$] and thus constitutes a valid local boundary perturbation that may be added to the Hamiltonian, at least when the intervening gapped domain is sufficiently small.  Moreover, since $\Theta$ abuts the two-particle-backscattering region, we can replace $\Theta \rightarrow \pi \hat n_\theta/2$, leaving
\begin{equation}
  \mathcal{O} \rightarrow \cos\left(\frac{\Phi}{2}-\frac{\pi}{2} \hat n_{\varphi}\right).
\end{equation}
The above operator has precisely the form of the perturbation in Eq.~\eqref{lambdapf}, which we now see indeed arises from hybridization with the parafermion zero mode as claimed in the main text.  It is also now apparent that such a perturbation preserves time-reversal symmetry as required; see Eqs.~\eqref{eq:time_reversal} and \eqref{Tns}.  This important property is not obvious in Eq.~\eqref{lambdapf} but becomes manifest in the explicit derivation presented here.

%%%%%%%%%%%%%%%%%%%%%%%%%%%%%%%%%%%%%%%%%%%%%%%%

\section{Perfect-crossed-Andreev-reflection fixed point for the short-superconductor setup}
\label{CAR}

Perfect crossed Andreev reflection represents a boundary condition for which an incoming electron from one side of the superconductor in Fig.~\ref{Device} converts into a co-moving hole at the other end.  Let $\psi_{R/L1} \sim e^{i(\Phi_1 \pm \Theta_1)}$ denote fermions at the left superconductor interface, and $\psi_{R/L2} \sim e^{i(\Phi_2 \pm \Theta_2)}$ denote fermions at the right interface.  Perfect crossed Andreev reflection implies the relations $\psi_{R1} = \psi_{R2}^\dagger$ and $\psi_{L1} = \psi_{L2}^\dagger$---which are clearly compatible with time-reversal symmetry, if present.  (More generally the electron and hole operators could differ by phase factors, which we ignore for simplicity.)  In terms of bosonized fields we get
\begin{equation}
  \Phi_1 = -\Phi_2 \equiv \Phi,~~~~~ \Theta_1 = -\Theta_2 \equiv \Theta.
  \label{CARdefinition}
\end{equation}
Integrating out fields away from the boundary yields the crossed-Andreev-reflection fixed point action
\begin{equation}
  S_{cA}[\Phi,\Theta] = \int \frac{d \omega}{2 \pi} \frac{|\omega|}{\pi} \left( g |\Phi_\omega|^2 + g^{-1} |\Theta_\omega|^2 \right),
\end{equation}
whose form is identical to Eq.~\eqref{TransmissionFixedPointAction}.

Consider the time-reversal-invariant situation.  Starting from either the Andreev$\oplus$Andreev or normal$\oplus$normal fixed points, which symmetry-preserving terms favor a flow toward perfect crossed Andreev reflection?  According to Eq.~\eqref{CARdefinition} such perturbations should favor pinning the sum $\Theta_1 + \Theta_2$ or $\Phi_1+\Phi_2$.  At the Andreev$\oplus$Andreev fixed point, the leading $\mathcal{T}$-preserving term that does this is $\propto\cos[2(\Theta_1 + \Theta_2)]$; recall Eq.~\eqref{eq:time_reversal}.  Crucially, this perturbation is less relevant than $\lambda_t$ defined in Eq.~\eqref{lambdat}, which drives a flow toward perfect normal transmission.  The leading perturbation at the normal$\oplus$normal fixed point that would favor perfect crossed Andreev reflection is $\propto \sin(\Phi_1+\Phi_2)$, which is again less relevant than the $\lambda_{e/2}$ term in Eq.~\eqref{NNperturbation} that favors perfect normal transmission.  Thus in both cases the onset of perfect crossed Andreev reflection seems highly unlikely.

If we explicitly break time-reversal symmetry, then the perfect-crossed-Andreev-reflection fixed point is in any case unstable for arbitrary $g \neq 1$---just as for perfect normal transmission, since the actions (and the leading physical perturbations) take exactly the same form. (Note that the marginal non-interacting $g=1$ case behaves quite differently. There, the low energy processes are dominated by normal transmission and crossed-Andreev reflection \cite{Nilsson_2008}.) Thus we are justified in considering only the limited set of fixed points discussed in Sec.~\ref{sec:superconductor}.

%%%%%%%%%%%%%%%%%%%%%%%%%%%%%%%%%%%%%%%%%%%%%%%%

\bibliography{QSHSC-references}

\end{document}